\documentclass[aps,prl,nobibnotes,twocolumn,superscriptaddress,bibliography]{revtex4-2}
\usepackage{amsfonts}
\usepackage{mathrsfs}
\usepackage{amsmath}
\usepackage{color}
\usepackage{natbib}
\usepackage{graphicx}
\usepackage{bm}
\usepackage{amssymb}
\usepackage{xspace}
\usepackage{epstopdf}
\usepackage{dcolumn}
\usepackage{longtable}
\usepackage{array}
\usepackage{makecell}
\usepackage{tabulary}
\usepackage{multirow}
\usepackage{appendix}
\usepackage{mathtools}
\usepackage{pifont}
\usepackage{longtable}
\newcolumntype{C}[1]{>{\centering\arraybackslash}p{#1}}
\newcolumntype{L}[1]{>{\flushleft\arraybackslash}p{#1}}

\usepackage{multirow}
\usepackage{epsfig}
\usepackage[normalem]{ulem}
\usepackage{amsmath}
\usepackage{braket}
\usepackage{bbold}
\usepackage{graphicx}

\makeatletter

\providecommand{\tabularnewline}{\\}

\usepackage[colorlinks=true, letterpaper=true, pdfstartview=FitV, linkcolor=blue, citecolor=blue, urlcolor=blue]{hyperref}

\makeatother
\begin{document}
\title{Type-II Antiferroelectricity}
\author{Yang Wang}
\affiliation{Key Lab of advanced optoelectronic quantum architecture and measurement (MOE), Beijing Key Lab of Nanophotonics $\&$ Ultrafine Optoelectronic Systems, and School of Physics, Beijing Institute of Technology, Beijing 100081, China}
\affiliation{International Center for Quantum Materials, Beijing Institute of Technology, Zhuhai 519000, China}

\author{Zhi-Ming Yu}
\email{zhiming\_yu@bit.edu.cn}
\affiliation{Key Lab of advanced optoelectronic quantum architecture and measurement (MOE), Beijing Key Lab of Nanophotonics $\&$ Ultrafine Optoelectronic Systems, and School of Physics, Beijing Institute of Technology, Beijing 100081, China}
\affiliation{International Center for Quantum Materials, Beijing Institute of Technology, Zhuhai 519000, China}

\author{Chaoxi Cui}
\affiliation{Key Lab of advanced optoelectronic quantum architecture and measurement (MOE), Beijing Key Lab of Nanophotonics $\&$ Ultrafine Optoelectronic Systems, and School of Physics, Beijing Institute of Technology, Beijing 100081, China}
\affiliation{International Center for Quantum Materials, Beijing Institute of Technology, Zhuhai 519000, China}

\author{Yilin Han}
\affiliation{Key Lab of advanced optoelectronic quantum architecture and measurement (MOE), Beijing Key Lab of Nanophotonics $\&$ Ultrafine Optoelectronic Systems, and School of Physics, Beijing Institute of Technology, Beijing 100081, China}
\affiliation{International Center for Quantum Materials, Beijing Institute of Technology, Zhuhai 519000, China}

\author{Tingli He}
\affiliation{Key Lab of advanced optoelectronic quantum architecture and measurement (MOE), Beijing Key Lab of Nanophotonics $\&$ Ultrafine Optoelectronic Systems, and School of Physics, Beijing Institute of Technology, Beijing 100081, China}
\affiliation{International Center for Quantum Materials, Beijing Institute of Technology, Zhuhai 519000, China}

\author{Weikang Wu}
\affiliation{Key Laboratory for Liquid-Solid Structural Evolution and Processing of Materials, Ministry of Education, Shandong University, Jinan 250061, China}

\author{Run-Wu Zhang}
\email{zhangrunwu@bit.edu.cn}
\affiliation{Key Lab of advanced optoelectronic quantum architecture and measurement (MOE), Beijing Key Lab of Nanophotonics $\&$ Ultrafine Optoelectronic Systems, and School of Physics, Beijing Institute of Technology, Beijing 100081, China}
\affiliation{International Center for Quantum Materials, Beijing Institute of Technology, Zhuhai 519000, China}

\author{Shengyuan A. Yang}
\email{shengyuan.yang@polyu.edu.hk}
\affiliation{Research Laboratory for Quantum Materials, Department of Applied Physics, The Hong Kong Polytechnic University, Kowloon, Hong Kong, China}

\author{Yugui Yao}
\email{ygyao@bit.edu.cn }
\affiliation{Key Lab of advanced optoelectronic quantum architecture and measurement (MOE), Beijing Key Lab of Nanophotonics $\&$ Ultrafine Optoelectronic Systems, and School of Physics, Beijing Institute of Technology, Beijing 100081, China}
\affiliation{International Center for Quantum Materials, Beijing Institute of Technology, Zhuhai 519000, China}

\begin{abstract}
Antiferroelectricity (AFE) is a fundamental concept in physics and materials science. Conventional AFEs have the picture of alternating local electric dipoles defined in real space. Here, we discover a new class of AFEs, termed type-II AFEs,
which possess opposite polarizations defined in momentum space across a pair of symmetry decoupled subspaces.
Unlike conventional AFEs, the order parameter of type-II AFEs is rigorously formulated through Berry-phase theory and can be quantitatively extracted from the electronic band structure. Focusing on a subclass of type-II AFEs that preserve spin-rotation symmetry, we establish the relevant symmetry constraints and identify all compatible spin point groups.
Remarkably, we find that type-II AFE order intrinsically coexists with antiferromagnetism, revealing a robust form of magnetoelectric coupling. We construct an altermagnetic model and identify several concrete antiferromagnetic/altermagnetic materials, such as FeS, Cr$_2$O$_3$, MgMnO$_3$, monolayer MoICl$_2$
and bilayer CrI$_3$, that exhibit this novel ordering. Furthermore, we uncover unique physical phenomena associated with type-II spin-AFE systems, including spin current generation upon AFE switching and localized spin polarization at boundaries and domain walls. Our findings reveal a previously hidden class of quantum materials with intertwined ferroic orders, offering exciting opportunities for both fundamental exploration and technological applications.
\end{abstract}

\maketitle

The concept of antiferroelectricity (AFE) was initially proposed in the 1950s~\cite{kittelTheoryAntiferroelectricCrystals1951a,shiraneDielectricPropertiesLead1951a,FerroelectricityversusAntiferroelectricity}. It refers to dielectric materials having spontaneously ordered electric dipoles arranged in an antiparallel manner, such that the net electric polarization vanishes.
Till now, AFEs have been discovered in several families of materials~\cite{Takezoe2010,Hao2014a,bennettOrthorhombicSemiconductorsAntiferroelectrics2013a}, and have attracted wide-spread interest due to their promising applications in high-performance capacitors, solid-state cooling, actuators, transducers, and memory devices~\cite{Principles,Chauhan2015,Yang2020,Si2024,Li2024a}.

\begin{figure}[t]
	\includegraphics[width=8 cm]{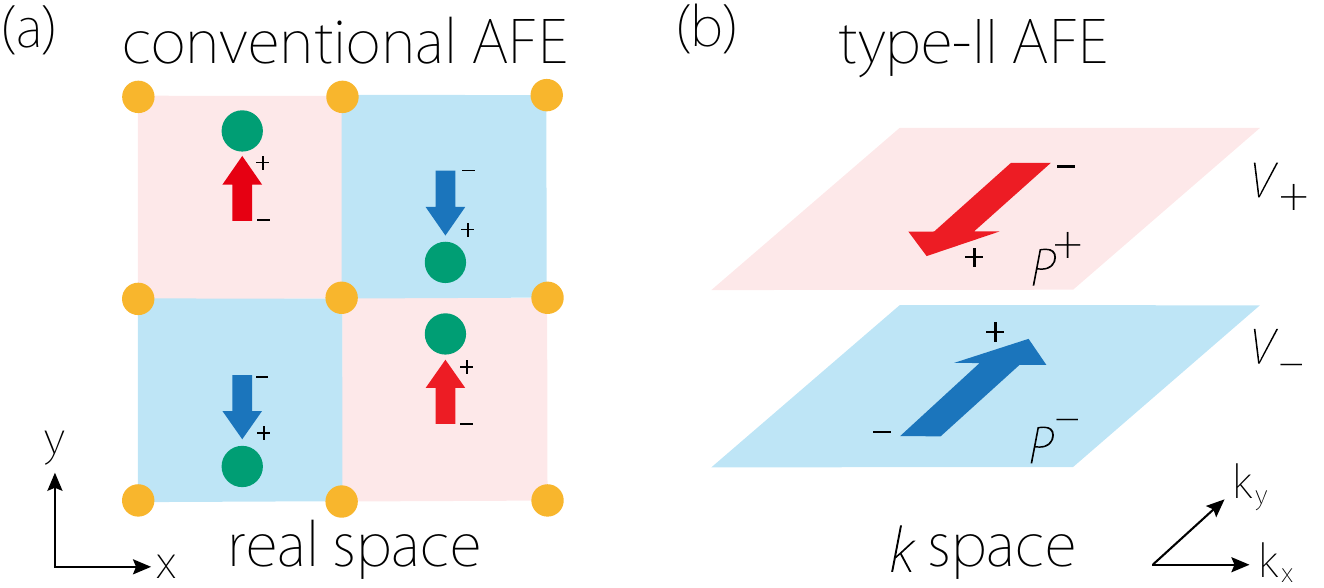}
	   \caption{(a) Illustration of conventional AFE. The green and yellow dots  represent two kinds of ions.
The AFE order is associated with opposite local electric dipoles defined in real space, as indicated by the red and blue arrows.
(b) Illustration of type-II AFE. The AFE order is connected to opposite polarization $P^\pm$ defined in $k$ space for two decoupled sectors $V_\pm$ of the total Hilbert space.}
	\label{fig1}
\end{figure}

The physical picture of these \emph{conventional} AFEs is a \emph{real-space} one, i.e., one needs to identify the \emph{local} electric dipoles as real-space quantities in the crystal lattice  as well as the antiparallel alignment of these dipoles [see Fig. \ref{fig1}(a)]. It is hence not fully compatible with band theory, which relies on Bloch states and a momentum-space description. This situation differs from
ferroelectricity, where its order parameter, the electric polarization, has a purely band theoretic formulation in terms of Berry phases of Bloch states~\cite{Resta1992,King-Smitha,Resta1994,vanderbiltBerryPhasesElectronic2018a}. In contrast, the AFE order parameter does not have a band formulation and can only be approximately estimated in practice~\cite{Xiao2018,Xiao2019,Xu2020,Liu2022}.

In this work, we propose a new class of AFE, which is termed as the type-II AFE. It refers to dielectrics which possess a symmetry allowing the decoupling of valence bands into two sectors. Each sector is polar and has a nonzero electric polarization, while polarizations of the two sectors are opposite, making a zero net polarization. For such type-II AFEs,
the order parameter has a well-defined Berry phase theory and can be quantitatively calculated from the band structure.
Focusing on the subclass of type-II AFEs with spin rotational symmetry, named as spin-AFE, we clarify its symmetry constraint and discover that AFE order must coexist with an antiferromagnetic (AFM) order, i.e., they are intrinsically multiferroic. Particularly, we show spin-AFE
can be realized in spin groups including several for altermagnets. We construct lattice models to demonstrate the essential physics and
identified several concrete spin-AFE materials, including $\mathrm{FeS}$, $\mathrm{Cr_{2}O_{3}}$, $\mathrm{MgMnO_{3}}$, monolayer $\mathrm{MoICl_{2}}$ and bilayer $\mathrm{CrI_{3}}$. These type-II AFEs may exhibit unique properties, such as
generation of pure spin current upon flip of AFE order, spin polarization at boundaries and domain walls, and strong
magnetoelectric coupling. Another subclass, i.e., type-II AFE with mirror symmetry, is also discussed.
Our work uncovers a new type of AFE materials, which host distinct characters from conventional AFEs.
Their multiferroic nature and unique properties are of great interest for fundamental research and potential device applications.

\global\long\def\arraystretch{1.4}%
\begin{table*}[t]
   \renewcommand{\arraystretch}{1.4}
	\caption{Collinear spin point groups for spin-AFE. The fifth column shows the type of magnetic order, and the last column indicates the direction of the AFE order parameter $\bm{\mathcal{Q}}=(\bm P^+-\bm P^-)/2$. AM denotes altermagnetism. Operations in $\mathcal G_0$ are spin-sector-preserving, while the operation $X$ switches the two spin sectors.
The direction of $\bm{\mathcal{Q}}$ is constrained by elements of $\mathcal G_0^R$ point group, e.g., a rotation forbids components of $\bm{\mathcal{Q}}$  in the plane perpendicular to the rotation axis, and a mirror forbids the component along the normal direction of the mirror plane.}
	\begin{ruledtabular} %
		\begin{tabular}{cllllllll}
			 Lattice  & $\mathcal G$ & Generators of $\mathcal G_0$  &  $\mathcal{G}_0^R$  &  $X$   & Magnetic order & $\bm{\mathcal{Q}}$ \tabularnewline
	\hline		
  Triclinic &
  ${}^{\overline{1}} \overline{1}{}^{\infty m} 1$  & $[C_2 T||E]$ &    $C_1$  &  $[C_2||\mathcal{P}]$ &  $\mathcal{P}T$-AFM& $(\mathcal{Q}_{1},\mathcal{Q}_{2},\mathcal{Q}_{3})$\tabularnewline
\cline{2-6}
  \multirow{2}{*}{Monoclinic}  &
  ${}^{1} 2/{}^{\overline{1}} m{}^{\infty m} 1$  & $[C_2 T||E],[E||2_{010}]$ &    $C_2$ & $[C_2||\mathcal{P}]$& $\mathcal{P}T$-AFM& $\mathcal{Q}_{2}$\tabularnewline
  &
 ${}^{\overline{1}} 2/{}^{1} m{}^{\infty m} 1$  & $[C_2 T||E],[E||m_{010}]$ &    $C_s$ & $[C_2||\mathcal{P}]$& $\mathcal{P}T$-AFM &  $(\mathcal{Q}_{1},0,\mathcal{Q}_{3})$ \tabularnewline
\cline{2-6}
\multirow{2}{*}{Orthorhombic}      &
 ${}^{1} 2{}^{\overline{1}} 2{}^{\overline{1}} 2{}^{\infty m} 1$  & $[C_2 T||E],[E||2_{001}]$ &    $C_2$ & $[C_2||2_{100}]$& AM  &$\mathcal{Q}_{3}$ \tabularnewline
   &
  ${}^{1} m{}^{1} m{}^{\overline{1}} m{}^{\infty m} 1$  & $[C_2 T||E],[E||m_{001}],[E||m_{010}]$ &    $C_{2v}$ & $[C_2||\mathcal{P}]$& $\mathcal{P}T$-AFM   &$\mathcal{Q}_{1}$ \tabularnewline
\cline{2-6}
 &
 ${}^{\overline{1}} \overline{4}{}^{\infty m} 1$  & $[C_2 T||E]$ &    $C_1$ &$[C_2||\overline{4}^{+}_{001}]$& AM &$\mathcal{Q}_{3}$ \tabularnewline
   &
  ${}^{1} 4/{}^{\overline{1}} m{}^{\infty m} 1$  & $[C_2T||E],[E||4^{+}_{001}]$ &   $C_4$ & $[C_2||\mathcal{P}]$& $\mathcal{P}T$-AFM  &$\mathcal{Q}_{3}$ \tabularnewline
    Tetragonal &
  ${}^{1} 4{}^{\overline{1}} 2{}^{\overline{1}} 2{}^{\infty m} 1$  & $[C_2T||E],[E||4^{+}_{001}]$ &    $C_4$ &$[C_2||2_{100}]$&  AM&$\mathcal{Q}_{3}$ \tabularnewline
 &  ${}^{\overline{1}} \overline{4}{}^{\overline{1}} 2{}^{1} m{}^{\infty m} 1$  & $[C_2T||E],[E||m_{110}],[E||2_{001}]$ &    $C_{2v}$ &$[C_2||\overline{4}^{+}_{001}]$& AM  & $\mathcal{Q}_{3}$ \tabularnewline
 & ${}^{1} 4/{}^{\overline{1}} m{}^{1} m{}^{1} m{}^{\infty m} 1$  & $[C_2T||E],[E||4^{+}_{001}],[E||m_{100}],[E||m_{010}]$ &    $C_{4v}$ &  $[C_2||\mathcal{P}]$ & $\mathcal{P}T$-AFM &$\mathcal{Q}_{3}$ \tabularnewline
\cline{2-6}
  &  ${}^{\overline{1}} \overline{3}{}^{\infty m} 1$  & $[C_2T||E]$ &    $C_1$ &   $[C_2||\mathcal{P}]$  & $\mathcal{P}T$-AFM &$\mathcal{Q}_{3}$ \tabularnewline
    Trigonal  & ${}^{1} 3{}^{\overline{1}} 2{}^{\infty m} 1$  & $[C_2T||E],[E||3^{+}_{001}]$ &    $C_3$ & $[C_2||2_{100}]$&AM  &$\mathcal{Q}_{3}$ \tabularnewline
    &  ${}^{\overline{1}} \overline{3}{}^{1} m{}^{\infty m} 1$  & $[C_2T||E],[E||m_{100}]$ &    $C_s$ & $[C_2||\mathcal{P}]$  & $\mathcal{P}T$-AFM &$\mathcal{Q}_{3}$ \tabularnewline
\cline{2-6}
    & ${}^{\overline{1}} \overline{6}{}^{\infty m} 1$  & $[C_2T||E]$ &    $C_1$ & $[C_2||\overline{6}^+_{001}]$& AM &$\mathcal{Q}_{3}$ \tabularnewline
   &  ${}^{1} 6{}^{\overline{1}} 2{}^{\overline{1}} 2{}^{\infty m} 1$  & $[C_2T||E],[E||6^{+}_{001}]$ &    $C_6$   &$[C_2||2_{100}]$&  AM&$\mathcal{Q}_{3}$ \tabularnewline
   Hexagonal  &  ${}^{1} 6/{}^{\overline{1}} m{}^{\infty m} 1$   & $[C_2T||E],[E||6^{+}_{001}]$ &    $C_6$ &$[C_2||\mathcal{P}]$& $\mathcal{P}T$-AFM  &$\mathcal{Q}_{3}$ \tabularnewline
    &  ${}^{\overline{1}} \overline{6}{}^{1} m{}^{\overline{1}} 2{}^{\infty m} 1$  & $[C_2T||E],[E||3^{+}_{001}],[E||m_{100}]$ &    $C_{3v}$ & $[C_2||\overline{6}^{+}_{001}]$  & AM& $\mathcal{Q}_{3}$\tabularnewline	
    & ${}^{1} 6/{}^{\overline{1}} m{}^{1} m{}^{1} m{}^{\infty m} 1$  & $[C_2T||E],[E||6^{+}_{001}],[E||m_{100}],[E||m_{1\overline{1}0}]$ &   $C_{6v}$ & $[C_2||\mathcal{P}]$& $\mathcal{P}T$-AFM &$\mathcal{Q}_{3}$ \tabularnewline
		\end{tabular}
	\end{ruledtabular}
	\label{table3}
\end{table*}

\textit{{Concept of type-II AFE.}}
Consider a band insulator with certain symmetry $W$ that decouples the Hilbert space $V$ of electronic states into two sectors (subspaces): $V=V_+\oplus V_-$. For example, if $W$ is the spin rotation symmetry, then $V_\pm$ would be the subspaces for the two spin species. Accordingly, the Bloch Hamiltonian $\mathcal H(\bm k)$ can also be decoupled into two blocks:
\begin{eqnarray}\label{eq:0}
\mathcal H(\bm{k})=h_{+}(\bm{k}) \oplus h_{-}(\bm{k}).
\end{eqnarray}

If the two subsystems defined by $h_\pm$ are polar, they each have a nonzero electric polarization, which can be evaluated by the Berry-phase theory~\cite{vanderbiltBerryPhasesElectronic2018a}, with
\begin{eqnarray}\label{eq:def}
{\bm P}^{\pm}=-e\sum_{n}^\text{occ}\int_\text{BZ}[d\bm k] {\bm{\mathcal A}}_n^{\pm}({\bm k}),
\end{eqnarray}
where $-e$ is the electron charge, the sum of band index $n$ is over valence bands, $[d\bm k]$ is a shorthand notation for $d^D k/(2\pi)^D$ with $D$ the dimension of the system, and ${\bm{\mathcal A}}_n^{\pm}(\bm k)=i\langle u_n^{\pm}({\bm k})| \nabla_{\bm k} |u_n^{\pm}({\bm k})\rangle$ is the Berry connection  with $|u_n^{\pm}({\bm k})\rangle$ the Bloch eigenstates of  subsystem  $h_{\pm}$. It should be noted that the polarization is well defined modulo $e\bm R/V_\text{cell}$~\cite{vanderbiltBerryPhasesElectronic2018a}, where $\bm R$ is a lattice vector and $V_\text{cell}$ is the unit cell volume.

Assume that the whole system $\mathcal{H}$ is nonpolar and there exists certain symmetry $X$ connecting $h_+$ and $h_-$, which ensures the net polarization vanishes:
${\bm P}={\bm P}^{+}+{\bm P}^{-}=0$.
Then, the system would be AFE, with a well-defined AFE order parameter from band theory:
\begin{eqnarray}\label{eq:def2}
\bm{\mathcal{{Q}}}&=&\frac{1}{2}({\bm P}^{+}-{\bm P}^{-}).
\end{eqnarray}

We have a few remarks here. First, as discussed above, for conventional AFEs, the evaluation of AFE order parameter relies on the identification of local dipoles in real space, which cannot be directly connected to band structure. In comparison, for type-II AFEs, the order parameter $\bm{\mathcal{{Q}}}$ permits a momentum-space formulation, and is determined as a band structure property.
This makes $\bm{\mathcal{{Q}}}$ well defined and readily evaluated from first-principles computations.

Second, the polarizations discussed above are from electronic contributions. Usually, there are also ionic contributions~\cite{vanderbiltBerryPhasesElectronic2018a}. However, as one can see below, by choosing the origin located at the center of $X$ symmetry operation, the ionic contribution can always be made zero.

Third, a key ingredient for type-II AFE is the decoupling of a nonpolar system into two polar subsystems under some  symmetry $W$. As mentioned, a natural case is with decoupling into two spin sectors under spin rotation symmetry, which is named as the spin-AFE. There also exist other possibilities. For example, two dimensional (2D) systems may have $W$ being horizontal mirror symmetry, and the decoupling is into the two subspaces with even and odd mirror eigenvalues. For 1D, one may also consider $W$ being twofold rotational symmetry along the axis of the system. In the following, we shall mainly focus on the spin-AFE subclass, which is the most interesting, and other cases shall be mentioned at the end.

\textit{{Symmetry condition for spin-AFE.}}
Spin-AFE occurs in systems with spin rotation symmetry, i.e., systems with negligible spin-orbit coupling (SOC).
It must be noted that the decoupling into two spin sectors does not require the full $SU(2)$ spin rotation group;
a $U(1)$ subgroup suffices. This extends scope of material platforms to also include the \emph{collinear} magnets, where the spin components $s_m$ in the magnetic moment direction are conserved and serve as the label of the two sectors.

First of all, one can easily see that spin-AFE cannot exist in nonmagnetic systems.
For a nonmagnetic system with spin rotation symmetry, the two spin sectors must have the same polarization, hence cannot realize AFE.
Besides, spin-AFE also cannot exist in ferromagnetic or ferrimagnetic systems: because the two spin sectors are inequivalent, their polarizations $\bm P^{\pm}$, if nonzero, cannot exactly satisfy 
$\bm P^+=-\bm P^-$.
It follows that the only possibility of spin-AFE is  in AFM systems.

Consider a collinear AFM system with negligible SOC. Let the local moments be in the direction $\hat{m}$, then the two spin sectors are labeled with $s_m=\pm$. The AFE order is constrained by the spin point group $\mathcal G$ of the system. It can always be decomposed as:
\begin{equation}
  \mathcal G=\mathcal G_0+X \mathcal G_0,
\end{equation}
where $\mathcal G_0$ is the set of symmetries preserving each spin sector, and $X\mathcal G_0$ are those switching the two sectors.

The elements of a spin point group is usually expressed in the form of $[S\| R]$, where $R$ is an ordinary point group element acting on real space, and $S$ acts independently on spin space~\cite{Litvin1977}.
For elements $[S\| R]\in \mathcal G_0$, we have $S\in \{E,C_2T\}\ltimes SO(2)$~\cite{Litvin1977}, where $E$ is the identity, $C_2$ is the spin-flip operation, and $T$ is time reversal; meanwhile, the collection of all the real-space operations $R$ forms a group, denoted as $\mathcal G_{0}^R$, which is the real-space  point group that preserves each spin sector. For each spin sector to have a nonzero polarization, we must have the condition (i) $\mathcal G_{0}^R$ must be a polar point group.

Furthermore, to have $\bm P^+=-\bm P^-$ exerts a requirement on symmetry $X$. Writing $X=[S_X\|R_X]$, this requirement translates into the condition (ii): The action of operation $R_X$ must reverse $\bm P^+$.

Based on these two conditions, we search through the 90 collinear spin point groups and find 18 groups that can accommodate spin-AFE. These candidate groups are listed in Table \ref{table3}. We observe that first, as expected, all these groups that can host spin-AFE correspond to systems also with AFM ordering. In other words, spin-AFE materials are intrinsically multiferroic, with both electric and magnetic orderings. Second, from our analysis above, magnetic ordering is the primary driving force for the emergence of spin AFE, which indicates a strong coupling between the two orderings and also justifies naming such systems as type-II AFEs. In addition, out of the 18 candidate groups in Table \ref{table3}, we find 10 groups are with $\mathcal{P}T$-symmetric AFM~\cite{mostovoyMultiferroicsDifferentRoutes2024}, whereas the other 8 are with altermagnetic ordering which is a currently an active research topic~\cite{Gonzalez-Hernandez2021,2022,Bai2024,PhysRevLett.133.056401,PhysRevLett.133.146602,Duan2025}. From Table \ref{table3}, such spin-AFE altermagnets can exist in a variety of lattice types, opening the possibility to
explore novel AFE altermagnets and interplay between AFE and altermagnetism.

\begin{figure}[t]
	\includegraphics[width=8.5 cm]{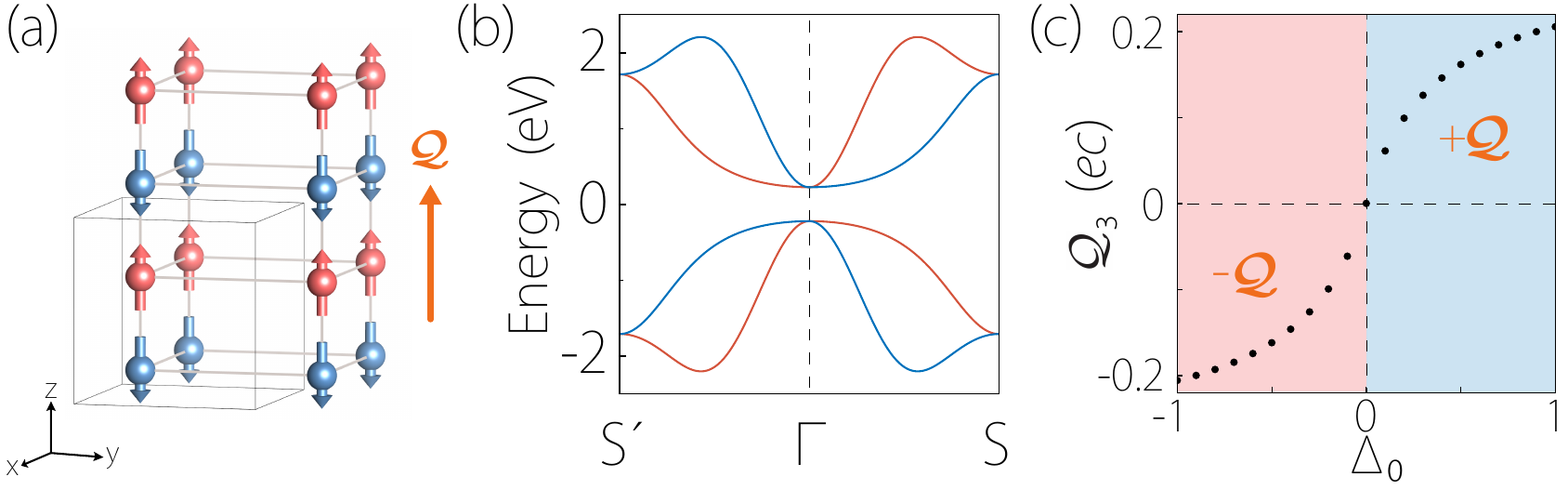}
	\caption{(a) Illustration of the spin-AFE lattice model (\ref{eq:ham1}). The red and blue arrows denote the magnetic moments on the sites.
(b) Band structure of the  model, showing altermagnetic spin splitting. The Brillouin zone is presented in Supplemental Material (SM) \cite{SM}.
The resulting AFE order $\bm{\mathcal Q}$ is along $z$ axis, as shown in (a).
(c) The AFE order  $\mathcal{Q}_3$ vs the AFM order $\Delta_0$.
	Here,  we set $t_{1}=-0.9$ eV, $t_{2} =0.8$ eV, and  $\chi=-0.8$ in (b-c) and  $\Delta_{0}=-1$ eV in (b).}
	\label{fig2}
\end{figure}

\textit{{Lattice model.}}
{We construct a simple  lattice model for spin-AFE with altermagnetism. Let's choose an  orthorhombic lattice belonging to spin point group ${}^{1} 2{}^{\overline{1}} 2{}^{\overline{1}} 2{}^{\infty m} 1$ (see Table~\ref{table3}).
Each unit of the lattice  contains two  sites  at positions $(0,0,\frac{1}{4})$ and $(0,0,\frac{3}{4})$.
The two sites each has two spin orbitals, and feature an AFM order, as illustrated in  Fig.~\ref{fig2}(a).
Then, a lattice Hamiltonian allowed by symmetry can be constructed as:
\begin{eqnarray}\label{eq:ham1}
	\mathcal{ H}(\bm{k})&=&
	\Delta_{0}\sigma_{3}\tau_{3}+t_+ \cos \frac{k_z}{2}\tau_{1}	+t_- \sin \frac{k_z}{2}\tau_{2}	\nonumber\\
	&&+	\chi \Delta_0(\sin k_x \sin k_y \sigma_{3}-\cos k_x \cos k_y \sigma_{3}\tau_{3}),
\end{eqnarray}
with $t_{\pm}(k_x)=\pm t_1+t_2 \cos k_x$, and   $\tau$'s and $\sigma$'s  are  Pauli matrices acting on site and spin spaces respectively, $\Delta_{0}$ is exchange term associated with AFM order, and $t$'s and $\chi$ are hopping parameters.

The band structure of model (\ref{eq:ham1}) is shown in Fig.~\ref{fig2}(b), showing the characteristic spin splitting for altermagnetism.
After computing the polarization of each spin subspace, we find that the spin-AFE order  $\bm{\mathcal{Q}}$ is along the $z$ axis, with ${\mathcal{Q}}_3=-0.206\ ec$ ($c$ is  the lattice constant).
Importantly, nonzero $\Delta_0$ is essential for finite spin-AFE order [see Fig. \ref{fig2}(c)], demonstrating that spin-AFE results from the AFM order. Moreover,
by  reversing the N\'eel  vector, the spin-AFE order $\bm{\mathcal{Q}}$ is also reversed [see Fig. \ref{fig2}(c)], showing  that spin-AFE materials are intrinsically multiferroic with strong magnetoelectric coupling.

}

\begin{figure}[t]
	\includegraphics[width=1\columnwidth]{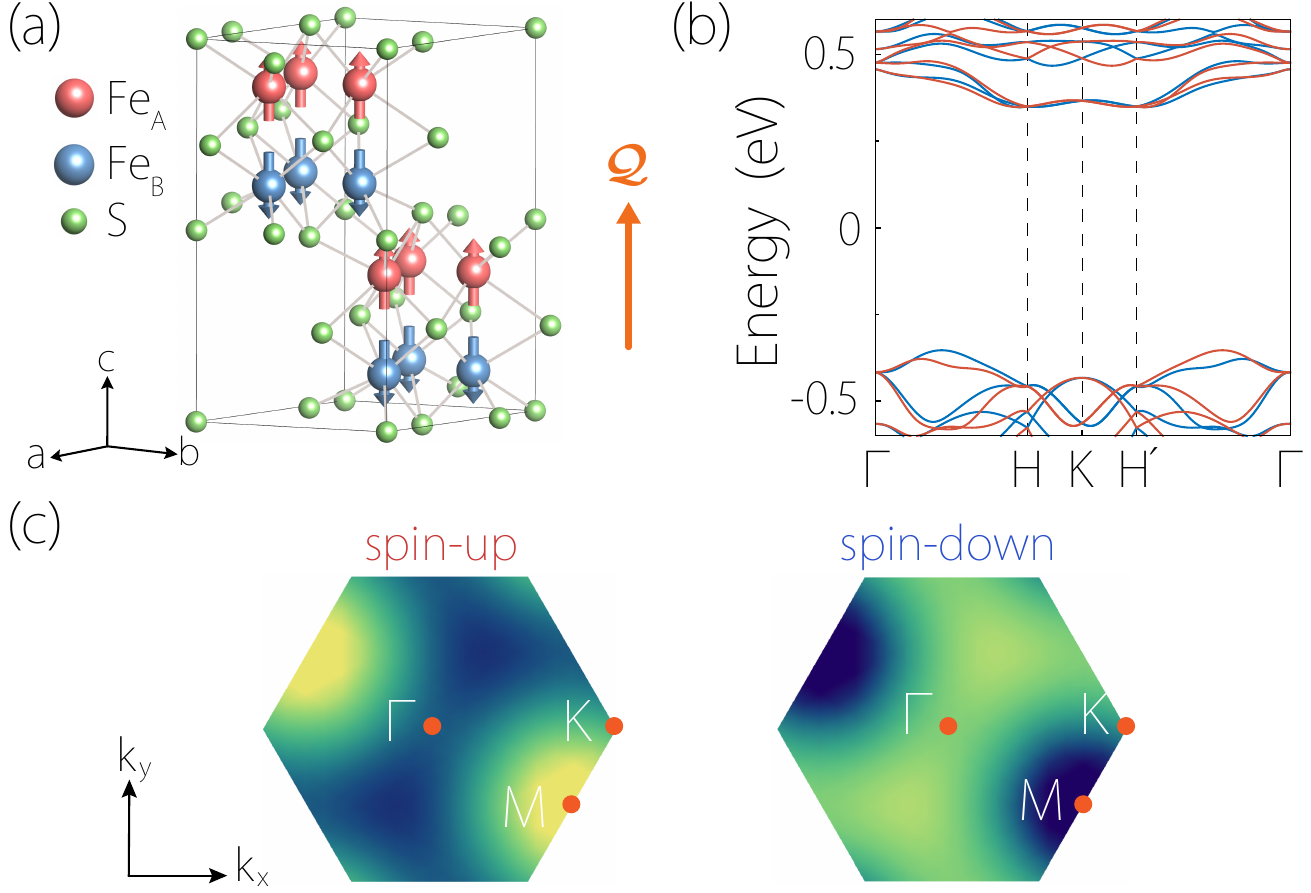}
	   \caption{(a) Crystalline  structure of  $\mathrm{FeS}$, where the red and blue spheres represent the $\mathrm{Fe}$ atoms with opposite magnetic moments. It is an altermagnetic type-II AFE.
(b) Band structure of $\mathrm{FeS}$ without SOC. Red and blue lines denote the spin-up and spin-down bands, showing altermagnetic spin splitting.
(c) The $k$ resolved polarization $P_z=\int_{0}^{2\pi}{\cal A}_z(k_x,k_y) dk_z$ in the $k_x$-$k_y$ plane. It takes opposite values for the two spin subspaces, leading to AFE.
}
\label{Fig3}
\end{figure}

\textit{{Material realization.}}
Guided by the symmetry conditions in Table \ref{table3}, we find several material candidates for spin-AFE. Here, we present two examples FeS and monolayer $\mathrm{MoICl_{2}}$. Other examples including $\mathrm{Cr_{2}O_{3}}$, $\mathrm{MgMnO_{3}}$ and bilayer $\mathrm{CrI_{3}}$ are given in Supplemental Materials (SM) \cite{SM}.

The first example FeS is a bulk collinear AFM, with a Morin transition temperature $\sim 220$ K~\cite{FeSScience,Knusn,Takagi2025}.
Experimentally, it was reported that FeS has a hexagonal crystal structure with lattice parameters $a=5.966$ \text{\AA} and $c=11.761$ \text{\AA} below 220 K~\cite{FeSScience,Knusn,Takagi2025}. The local moments are mainly on the Fe sites, with an easy axis in the $c$ direction forming a collinear AFM configuration, as illustrated in Fig. \ref{Fig3}(a).
In the AFM state, the spin point group is ${}^{\overline{1}} \overline{6}{}^{1} m{}^{\overline{1}} 2{}^{\infty m} 1$~\cite{SM}. According to Table \ref{table3}, the system is an altermagnet. Its $\mathcal G_0^R$ group is $C_{3v}$, which, for each spin sector, allows a polarization in the $z$ direction ($c$ axis). And its $X$ operation $[C_2 \|\bar 6^+_{001}]$ guarantees the two spin sectors have opposite $P_z$.
Therefore, FeS should have an AFE order $\bm{\mathcal{Q}}$ along $z$ direction.

In Fig.~\ref{Fig3}(b), we plot the calculated band structure of FeS in the absence of SOC. One observes it is  a magnetic semiconductor with a band gap $\sim 0.64$ eV. The spin splitting in the band structure can be clearly seen, manifesting its altermagnetic character.
We also checked that SOC only has weak effect on the band structure \cite{SM}.
Using Eqs.~(\ref{eq:def}) and (\ref{eq:def2}), we compute the AFE order parameter $\bm{\mathcal{Q}}$. For type-II AFEs, since the lattice is nonpolar, $\mathcal Q$ is solely from the electronic contribution, computed by the Berry-phase method for each spin sector.  The result confirms that $\bm{\mathcal{Q}}$ is along $z$ direction, with a value $\sim 1.550$ $\mathrm{\mu C/cm^2}$.
This value is comparable to the ferroelectric polarization in perovskite oxides~\cite{FunctionalMetalOxides}.
In addition, consistent with observation on model (\ref{eq:ham1}), we find that the type-II AFE order vanishes above magnetic transition, and is flipped under the reversal of N\'eel vector.
These results reveal  FeS as an altermagnetic spin-AFE with sizable AFE and strong coupling between AFM and AFE orders.

\begin{figure}[t]
	\centering
	\includegraphics[width=1\columnwidth]{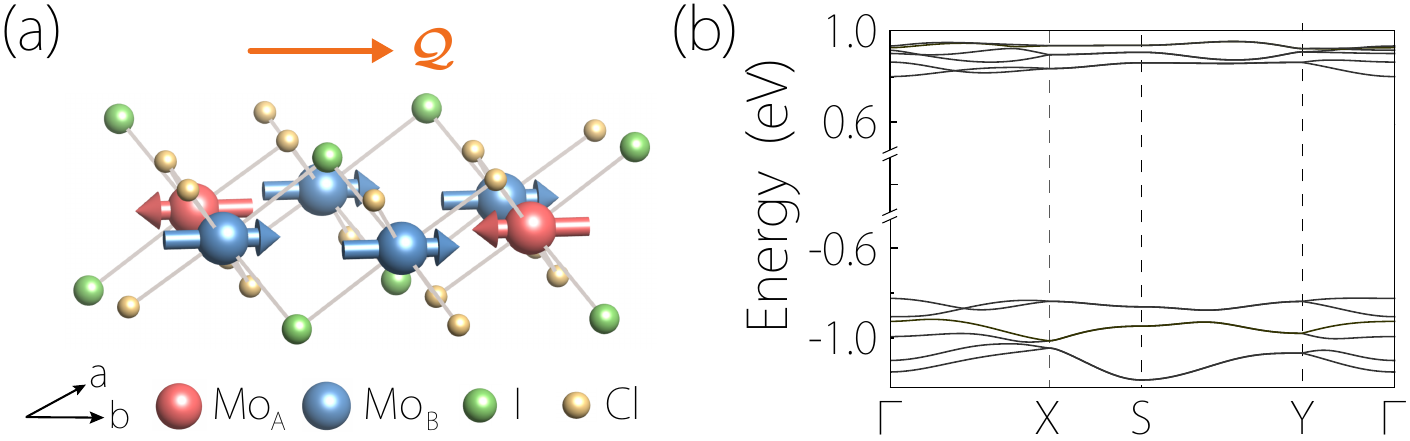}
	\caption{(a) Crystalline structure of monolayer $\mathrm{MoICl_2}$, where the red and blue spheres represent $\mathrm{Mo}$ atoms with opposite magnetic moments. Its AFE order $\bm{\mathcal Q}$ is along $b$ axis.
		(b) Band structure of monolayer $\mathrm{MoICl_2}$ without SOC, which are spin-degenerate.
	}
	\label{fig4}
\end{figure}

As the second example, we consider a 2D system: monolayer $\mathrm{MoICl_{2}}$~\cite{Lan2023}.
It has the $\mathrm{CrI}_{3}$-type structure, consisting of three atomic layers $\mathrm{Cl}(\mathrm{I})$-$\mathrm{Mo}$-$\mathrm{Cl}(\mathrm{I})$, as shown in Fig. \ref{fig4}(a). The lattice constants are $a=6.63$ \text{\AA} and $b=11.48$ \text{\AA} \cite{SM}.
The ground state of $\mathrm{MoICl_{2}}$ is AFM, with the local spin configuration in Fig. \ref{fig4}(a). The magnetic easy axis is along $(1,0)$ direction. This state has spin layer-point group  $[{}^{1} 2/{}^{\overline{1}} m{}^{\infty m} 1]_2$~\cite{SM}, corresponding to a $\mathcal{P}T$-symmetric AFM.
From Table \ref{table3}, its AFE order $\bm{\mathcal{Q}}$ should be along $b$ axis. Figure \ref{fig4}(b) shows the calculated band structure in the absence of SOC, from which $\mathcal{Q}_2$ is found to be 11.875 pC/m, consistent with our expectation. This value is also quite sizable compared to other 2D materials such as $\mathrm{WTe_{2}}$~\cite{Fei2018}.

\textit{{Discussion.}}
We have proposed a new class of materials with AFE ordering. Distinct from conventional AFE materials,
these type-II AFEs are intrinsically multiferroic: their AFE orders must coexist and are actually resulted from
magnetic AFM ordering.  This necessarily indicates a strong magnetoelectric coupling: reversing one will also reverse the other.
We have demonstrated that
flipping the magnetic N\'eel vector, e.g., by electric N\'eel torque or optical method \cite{Baltz2018b,Nemec2018,Manchon2019a}, will reverse the AFE order.
One can expect that applied magnetic field, via spin-flop or spin-flip transitions~\cite{Huang2018a,Jiang2018,Amin2024,He2024a}, can also strongly influence the
AFE order. Compared to conventional AFEs, another important difference of type-II AFEs is on the hysteresis behavior under electric field. For conventional AFEs, the electric dipoles are defined in real space, often associated with some local lattice distortions, e.g., displacement of ions in perovskite structure. Hence, usually one can identify a metastable
ferroelectric state with all local dipoles aligned in the same direction, and the transitions between AFE and this ferroelectric state lead to a double hysteresis loop often observed in conventional AFEs~\cite{Takezoe2010,Hao2014a}.
In comparison, for type-II AFEs,
there is no simple identification of a metastable ferroelectric state (it depends on details of specific material), so a type-II AFE may or may not exhibit double hysteresis loops. For FeS, we identify such a ferroelectric state~\cite{SM}, and a double hysteresis loop may be observed in practice.

\begin{figure}[t]
	\includegraphics[width=1\columnwidth]{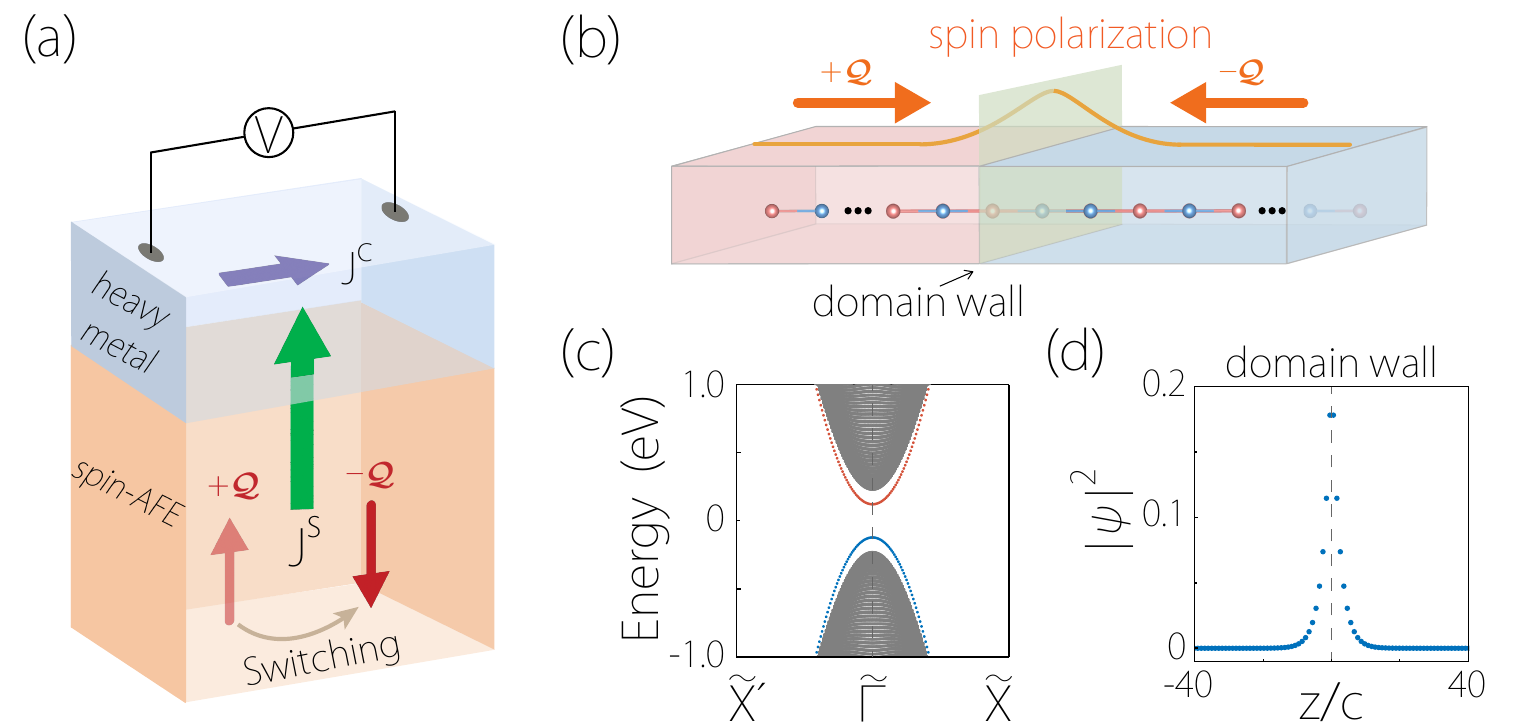}
	   \caption{(a) Schematic  of a possible  junction geometry for detection of spin-AFE. A pure spin current is generated by reversing the spin-AFE order, and it flows into an adjacent heavy-metal layer and produces a transverse voltage signal via the inverse spin Hall effect.
(b) Domain wall between two spin-AFE domains hosts a local spin polarization.
(c) Calculated band structure for such a domain wall based on model (\ref{eq:ham1}).
The red and blue dots are respectively the spin-up and spin-down domain wall modes.
(d) The real-space distribution of spin-down domain wall mode at $\Gamma$ point in (c).}
	\label{fig5}
\end{figure}

Type-II AFE also manifest other interesting properties. For example, in ferroelectrics, the change of polarization induces a charge current~\cite{vanderbiltBerryPhasesElectronic2018a}. In analogy, in spin-AFEs, the change of AFE order parameter will generate a pure spin current:
\begin{equation}
  j^p_a= \frac{\partial}{\partial t}\mathcal Q_a,
\end{equation}
where polarization $p$ of spin current $j^p_a$ is the N\'eel vector direction, and $a$ labels the Cartesian component.
Consider the material $\mathrm{FeS}$, the reverse of its AFE order will generate a spin current in the $c$ direction. Assuming the reversal process occurs in a time scale of ps (which is the typical time scale for switching antiferromagnets), the generated spin current can reach a magnitude of order $10^{10}$ $\mathrm{A/m^2}$. 
This pure spin current can be detected by inverse spin Hall effect after injecting it into a nearby metallic layer [see Fig. \ref{fig5}(a)], or by its induced magnetization dynamics after injecting it into a nearly ferromagnet~\cite{Hellman2017,Manchon2019a}.

In addition, different from conventional AFEs, spin-AFE could host spin polarization at system boundaries, domain walls, and other topological defects. For example, at a boundary, we should have spin polarization density $s^p=\bm{\mathcal Q}\cdot \hat n$ where $\hat n$ is the boundary normal vector.  Figure \ref{fig5}(c) shows the calculated electronic spectrum of a head-to-head AFE domain wall for model (\ref{eq:ham1}) [see Fig. \ref{fig5}(b)].  One observes two bands inside the band gap. They each is localized at the domain wall, corresponding to a polarized spin density [see Fig. \ref{fig5}(d)]. Such local spin density can be probed by magneto-optical measurement~\cite{Science2004}.

Finally, we mentioned there are other subclasses of type-II AFEs, such as those enabled by mirror symmetry or twofold rotation in reduced dimensions. In SM~\cite{SM}, we give a symmetry analysis for the mirror type-II AFEs. We find such AFEs must also coexist with AFM ordering, and some material candidates are predicted.
The search for more type-II AFE systems and the study of their unique physical properties will be interesting directions for future research.

\acknowledgements
We thank Zeying Zhang and D. L. Deng for helpful discussions. This work is supported by the National Natural Science Foundation of China (Grants No. 12234003, No. 12474040, No. 12304188, No. 12321004),  HK PolyU start-up fund (P0057929), the Natural Science Foundation of Beijing (Grant No. 1252029), the Shandong Provincial Natural Science Foundation (No. ZR2023QA012), and the program of Outstanding Young and Middle-aged Scholars of Shandong University.

\textit{Data availability}--The data that support the findings of this Letter are openly available~\cite{DataZenodo}.

\bibliography{ref}

\end{document}


\title{Supplemental material for ``Type-II Antiferroelectricity"}
\author{Yang Wang}
\affiliation{Key Lab of advanced optoelectronic quantum architecture and measurement (MOE), Beijing Key Lab of Nanophotonics $\&$ Ultrafine Optoelectronic Systems, and School of Physics, Beijing Institute of Technology, Beijing 100081, China}
\affiliation{International Center for Quantum Materials, Beijing Institute of Technology, Zhuhai 519000, China}

\author{Zhi-Ming Yu}
\affiliation{Key Lab of advanced optoelectronic quantum architecture and measurement (MOE), Beijing Key Lab of Nanophotonics $\&$ Ultrafine Optoelectronic Systems, and School of Physics, Beijing Institute of Technology, Beijing 100081, China}
\affiliation{International Center for Quantum Materials, Beijing Institute of Technology, Zhuhai 519000, China}

\author{Chaoxi Cui}
\affiliation{Key Lab of advanced optoelectronic quantum architecture and measurement (MOE), Beijing Key Lab of Nanophotonics $\&$ Ultrafine Optoelectronic Systems, and School of Physics, Beijing Institute of Technology, Beijing 100081, China}
\affiliation{International Center for Quantum Materials, Beijing Institute of Technology, Zhuhai 519000, China}

\author{Yilin Han}
\affiliation{Key Lab of advanced optoelectronic quantum architecture and measurement (MOE), Beijing Key Lab of Nanophotonics $\&$ Ultrafine Optoelectronic Systems, and School of Physics, Beijing Institute of Technology, Beijing 100081, China}
\affiliation{International Center for Quantum Materials, Beijing Institute of Technology, Zhuhai 519000, China}

\author{Tingli He}
\affiliation{Key Lab of advanced optoelectronic quantum architecture and measurement (MOE), Beijing Key Lab of Nanophotonics $\&$ Ultrafine Optoelectronic Systems, and School of Physics, Beijing Institute of Technology, Beijing 100081, China}
\affiliation{International Center for Quantum Materials, Beijing Institute of Technology, Zhuhai 519000, China}

\author{Weikang Wu}
\affiliation{Key Laboratory for Liquid-Solid Structural Evolution and Processing of Materials, Ministry of Education, Shandong University, Jinan 250061, China}

\author{Run-Wu Zhang}
\affiliation{Key Lab of advanced optoelectronic quantum architecture and measurement (MOE), Beijing Key Lab of Nanophotonics $\&$ Ultrafine Optoelectronic Systems, and School of Physics, Beijing Institute of Technology, Beijing 100081, China}
\affiliation{International Center for Quantum Materials, Beijing Institute of Technology, Zhuhai 519000, China}

\author{Shengyuan A. Yang}
\affiliation{Research Laboratory for Quantum Materials, Department of Applied Physics, The Hong Kong Polytechnic University, Kowloon, Hong Kong, China}

\author{Yugui Yao}
\affiliation{Key Lab of advanced optoelectronic quantum architecture and measurement (MOE), Beijing Key Lab of Nanophotonics $\&$ Ultrafine Optoelectronic Systems, and School of Physics, Beijing Institute of Technology, Beijing 100081, China}
\affiliation{International Center for Quantum Materials, Beijing Institute of Technology, Zhuhai 519000, China}

\maketitle

\tableofcontents
\newpage

\section{Methods}
In this work, all first-principles calculations are performed using density functional theory (DFT) within the Vienna Ab initio Simulation Package (VASP)~\cite{PRB11169}, employing the projector augmented wave (PAW) method~\cite{2}. The exchange-correlation interaction is calculated by the generalized gradient approximation with the Perdew-Burke-Ernzerhof realization~\cite{1}. For calculations expect for polarization calculations, the plane-wave basis set is defined by a kinetic energy cutoff of at least 550 eV, while Brillouin zone integration employed a $\Gamma$-centered Monkhorst-Pack \textit{k}-mesh with a space of at least $2\pi \times 0.03$. The energy and force convergence criteria are set to be not lower than $10^{-6}$ eV and $10^{-2}$ eV/\AA, respectively. A vacuum space (20 \AA) for two-dimensional (2D) systems is introduced to avoid interactions between neighboring slabs. The DFT+\textit{U} method~\cite{3} which corrects electron self-interaction was used for all calculations. Herein, we provide specific \textit{U} values for transition-metal elements and details for polarization calculation in the following different sections. 

\section{Brillouin zones for models and materials in main text}

\begin{figure}[H]
	\centering
	\includegraphics[width=0.6\columnwidth]{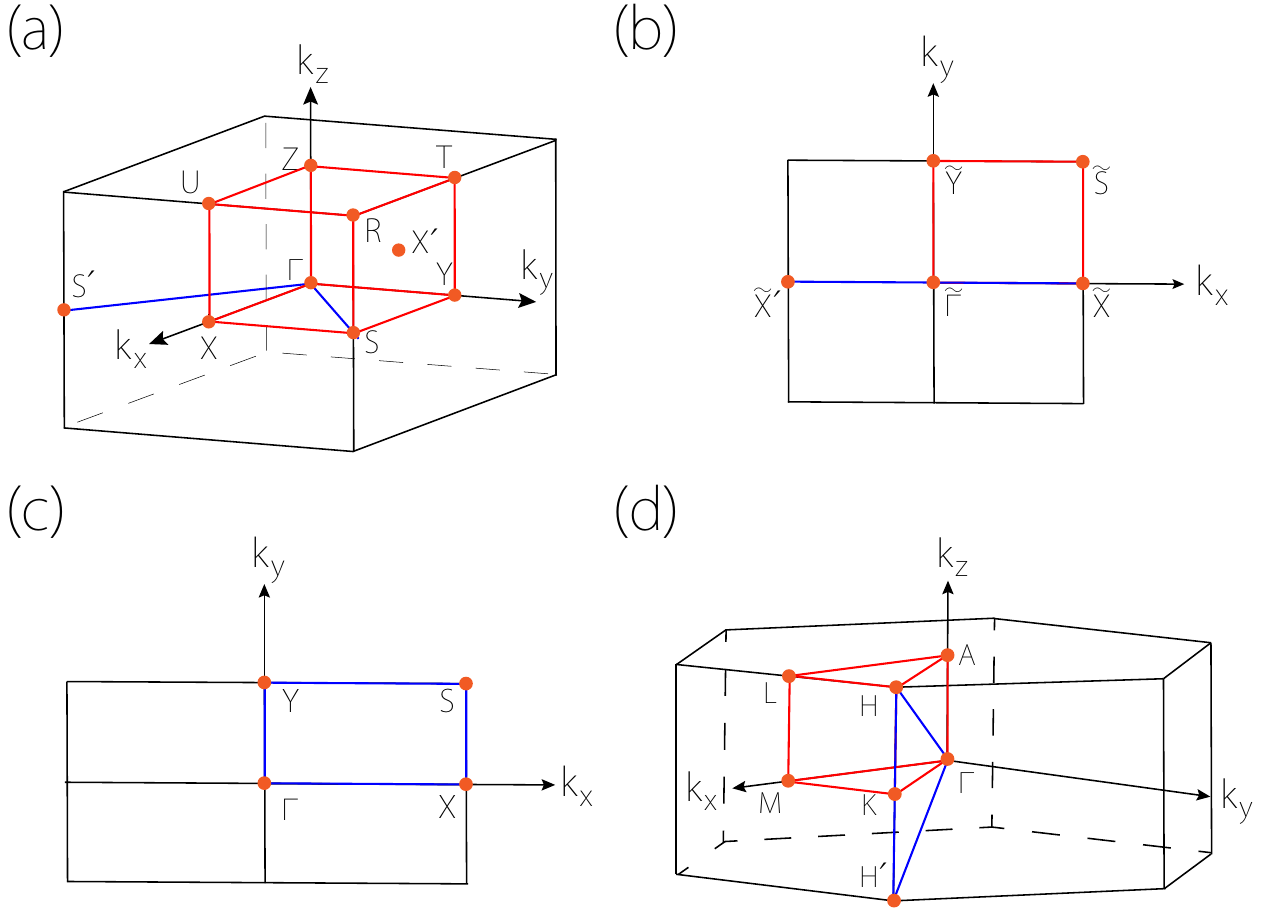}
	\caption{(a) BZ of lattice model (5) in the main text. (b) BZ of junction model illustrated in Fig. 5(b) in the main text. BZs of (c)  monolayer $\mathrm{MoICl_2}$ and (d) FeS.}
	\label{figs12}
\end{figure}

Figure \ref{figs12} shows  the Brillouin zones (BZs)  of  the models and materials discussed in the main text. 
Specifically, the BZ of the lattice model (5) in the main text is presented in Fig. \ref{figs12}(a), and the BZ of the junction model [illustrated in Fig. 5(b) in the main text] is presented in Fig. \ref{figs12}(b).
The BZs of  monolayer $\mathrm{MoICl_2}$ and FeS are shown as Fig. \ref{figs12}(c) and Fig. \ref{figs12}(d), respectively.

\section{Collinear spin layer-point groups for 2D spin-AFE}

In the main text, we provide a detailed symmetry analysis for the 3D spin-AFE. 
The spin-AFE can also exist in 2D systems with negligible SOC. 
Similarly, we search through the 94 collinear spin layer-point groups (which describe the symmetry of the 2D systems that has a  finite  thickness along out-of-plane direction) and find 22 groups that can accommodate  2D spin-AFE. These candidate groups are
listed in Table \ref{table2DspinAFE}.

\global\long\def\arraystretch{1.5}%
\begin{table}[H]
	\caption{Collinear spin layer-point groups for 2D spin-AFE. The fifth column shows the type of magnetic order, and the last column indicates the direction of the AFE order parameter $\bm{\mathcal{Q}}$. The subscript $i$ in the notation of $\mathcal G$: $[...]_i$ indicates the $i$-th collinear spin layer-point group obtained from a same 3D collinear spin point group.  }
	\begin{ruledtabular} %
		\begin{tabular}{cllllllll}
			Lattice  & $\mathcal G$ & Generators of $\mathcal G_0$  & Symmetry $X$   & Magnetic order & $\bm{\mathcal{Q}}$ \tabularnewline
			\hline		
			Triclinic &
			${}^{\overline{1}} \overline{1}{}^{\infty m} 1$  & $[C_2 T||E]$ &$[C_2||\mathcal{P}]$ &$\mathcal{P}T$-AFM& $(\mathcal{Q}_{1},\mathcal{Q}_{2},\mathcal{Q}_{3})$\tabularnewline
			\cline{2-6}
			\multirow{4}{*}{Monoclinic}  &
			$[{}^{1} 2/{}^{\overline{1}} m{}^{\infty m} 1]_1$  & $[C_2 T||E],[E||2_{001}]$ & $[C_2||\mathcal{P}]$& $\mathcal{P}T$-AFM& $\mathcal{Q}_{3}$\tabularnewline
			&
			$[{}^{1} 2/{}^{\overline{1}} m{}^{\infty m} 1]_2$  & $[C_2 T||E],[E||2_{100}]$ & $[C_2||\mathcal{P}]$& $\mathcal{P}T$-AFM& $\mathcal{Q}_{1}$\tabularnewline
			&
			$[{}^{\overline{1}} 2/{}^{1} m{}^{\infty m} 1]_1$  & $[C_2 T||E],[E||m_{001}]$ & $[C_2||\mathcal{P}]$& $\mathcal{P}T$-AFM &  $(\mathcal{Q}_{1},\mathcal{Q}_{2},0)$ \tabularnewline
			&
			$[{}^{\overline{1}} 2/{}^{1} m{}^{\infty m} 1]_2$  & $[C_2 T||E],[E||m_{100}]$ & $[C_2||\mathcal{P}]$& $\mathcal{P}T$-AFM &  $(0,\mathcal{Q}_{2},\mathcal{Q}_{3})$ \tabularnewline
			\cline{2-6}
			\multirow{4}{*}{Orthorhombic}      &
			$[{}^{1} 2{}^{\overline{1}} 2{}^{\overline{1}} 2{}^{\infty m} 1]_1$  & $[C_2 T||E],[E||2_{001}]$ & $[C_2||2_{100}]$& AM  &$\mathcal{Q}_{3}$ \tabularnewline
			&
			$[{}^{1} 2{}^{\overline{1}} 2{}^{\overline{1}} 2{}^{\infty m} 1]_2$  & $[C_2 T||E],[E||2_{010}]$ & $[C_2||2_{100}]$& AM  &$\mathcal{Q}_{2}$ \tabularnewline
			&
			$[{}^{1} m{}^{1} m{}^{\overline{1}} m{}^{\infty m} 1]_1$  & $[C_2 T||E],[E||m_{001}],[E||m_{010}]$ & $[C_2||\mathcal{P}]$& $\mathcal{P}T$-AFM   &$\mathcal{Q}_{1}$ \tabularnewline
			&
			$[{}^{1} m{}^{1} m{}^{\overline{1}} m{}^{\infty m} 1]_2$  & $[C_2 T||E],[E||m_{100}],[E||m_{010}]$ & $[C_2||\mathcal{P}]$& $\mathcal{P}T$-AFM   &$\mathcal{Q}_{3}$ \tabularnewline
			\cline{2-6}
			&
			${}^{\overline{1}} \overline{4}{}^{\infty m} 1$  & $[C_2 T||E]$ &$[C_2||\overline{4}^{+}_{001}]$& AM &$\mathcal{Q}_{3}$ \tabularnewline
			&
			${}^{1} 4/{}^{\overline{1}} m{}^{\infty m} 1$  & $[C_2T||E],[E||4^{+}_{001}]$ & $[C_2||\mathcal{P}]$& $\mathcal{P}T$-AFM  &$\mathcal{Q}_{3}$ \tabularnewline
			Tetragonal &
			${}^{1} 4{}^{\overline{1}} 2{}^{\overline{1}} 2{}^{\infty m} 1$  & $[C_2T||E],[E||4^{+}_{001}]$ &$[C_2||2_{100}]$&  AM&$\mathcal{Q}_{3}$ \tabularnewline
			&  ${}^{\overline{1}} \overline{4}{}^{\overline{1}} 2{}^{1} m{}^{\infty m} 1$  & $[C_2T||E],[E||m_{110}],[E||2_{001}]$ &$[C_2||\overline{4}^{+}_{001}]$& AM  & $\mathcal{Q}_{3}$ \tabularnewline
			& ${}^{1} 4/{}^{\overline{1}} m{}^{1} m{}^{1} m{}^{\infty m} 1$  & $[C_2T||E],[E||4^{+}_{001}],[E||m_{100}],[E||m_{010}]$ &  $[C_2||\mathcal{P}]$ & $\mathcal{P}T$-AFM &$\mathcal{Q}_{3}$ \tabularnewline
			\cline{2-6}
			&  ${}^{\overline{1}} \overline{3}{}^{\infty m} 1$  & $[C_2T||E]$ &   $[C_2||\mathcal{P}]$  & $\mathcal{P}T$-AFM &$\mathcal{Q}_{3}$ \tabularnewline
			Trigonal  & ${}^{1} 3{}^{\overline{1}} 2{}^{\infty m} 1$  & $[C_2T||E],[E||3^{+}_{001}]$ & $[C_2||2_{100}]$&AM  &$\mathcal{Q}_{3}$ \tabularnewline
			&  ${}^{\overline{1}} \overline{3}{}^{1} m{}^{\infty m} 1$  & $[C_2T||E],[E||m_{100}]$ &  $[C_2||\mathcal{P}]$  & $\mathcal{P}T$-AFM &$\mathcal{Q}_{3}$ \tabularnewline
			\cline{2-6}
			& ${}^{\overline{1}} \overline{6}{}^{\infty m} 1$  & $[C_2T||E]$ & $[C_2||\overline{6}^+_{001}]$& AM &$\mathcal{Q}_{3}$ \tabularnewline
			&  ${}^{1} 6{}^{\overline{1}} 2{}^{\overline{1}} 2{}^{\infty m} 1$  & $[C_2T||E],[E||6^{+}_{001}]$ &$[C_2||2_{100}]$&  AM&$\mathcal{Q}_{3}$ \tabularnewline
			Hexagonal  &  ${}^{1} 6/{}^{\overline{1}} m{}^{\infty m} 1$  & $[C_2T||E],[E||6^{+}_{001}]$ &$[C_2||\mathcal{P}]$& $\mathcal{P}T$-AFM  &$\mathcal{Q}_{3}$ \tabularnewline
			&  ${}^{\overline{1}} \overline{6}{}^{1} m{}^{\overline{1}} 2{}^{\infty m} 1$  & $[C_2T||E],[E||3^{+}_{001}],[E||m_{100}]$ & $[C_2||\overline{6}^{+}_{001}]$& AM& $\mathcal{Q}_{3}$\tabularnewline	
			& ${}^{1} 6/{}^{\overline{1}} m{}^{1} m{}^{1} m{}^{\infty m} 1$  & $[C_2T||E],[E||6^{+}_{001}],[E||m_{100}],[E||m_{1\overline{1}0}]$ & $[C_2||\mathcal{P}]$& $\mathcal{P}T$-AFM &$\mathcal{Q}_{3}$ \tabularnewline
		\end{tabular}
	\end{ruledtabular}
	\label{table2DspinAFE}
\end{table}

\section{Table of space groups of the materials}

In Table \ref{tablesadd3}, we present the space groups of  the  materials studied here.

\begin{table}[!htbp]
	\renewcommand{\arraystretch}{1.2}
	\caption{The space groups of the  materials studied here.}
	\begin{ruledtabular} %
		\begin{tabular}{ccccccc}
			$\mathrm{FeS}$ & Monolayer $\mathrm{MoICl_2}$ & $\mathrm{Cr_2O_3}$ & $\mathrm{MgMnO_3}$ & Bilayer $ \mathrm{CrI_3}$ & Monolayer $\mathrm{Sr_3X_2Cl_2O_5}$ \tabularnewline
			\hline
$P\overline{6}2c$ (No. 190) &  $C2/m$ (No.12) & $R\overline{3}c$ (No.167) & $R\overline{3}$ (No.148) & $C2/m$ (No.12) & $P4/mmm$ (No.123) \tabularnewline

		\end{tabular}
	\end{ruledtabular}
	\label{tablesadd3}
\end{table}

\section{Material candidates for  spin-AFE}
Guided by the symmetry conditions in Table I in the main text and Table \ref{table2DspinAFE}, we find several material candidates for spin-AFE, including FeS, Cr$_2$O$_3$, MgMnO$_3$, monolayer MoICl$_2$
and bilayer CrI$_3$. Among them, FeS and monolayer MoICl$_2$ are presented in the main text. Here, we introduce Cr$_2$O$_3$, MgMnO$_3$, bilayer CrI$_3$ and the calculation details of FeS and monolayer MoICl$_2$.

\subsection{3D systems} 
\subsubsection{$\mathcal{P}T$-AFM material $\mathrm{Cr_{2}O_{3}}$}

\begin{figure}[htbp]
	\centering
	\includegraphics[width=0.7\columnwidth]{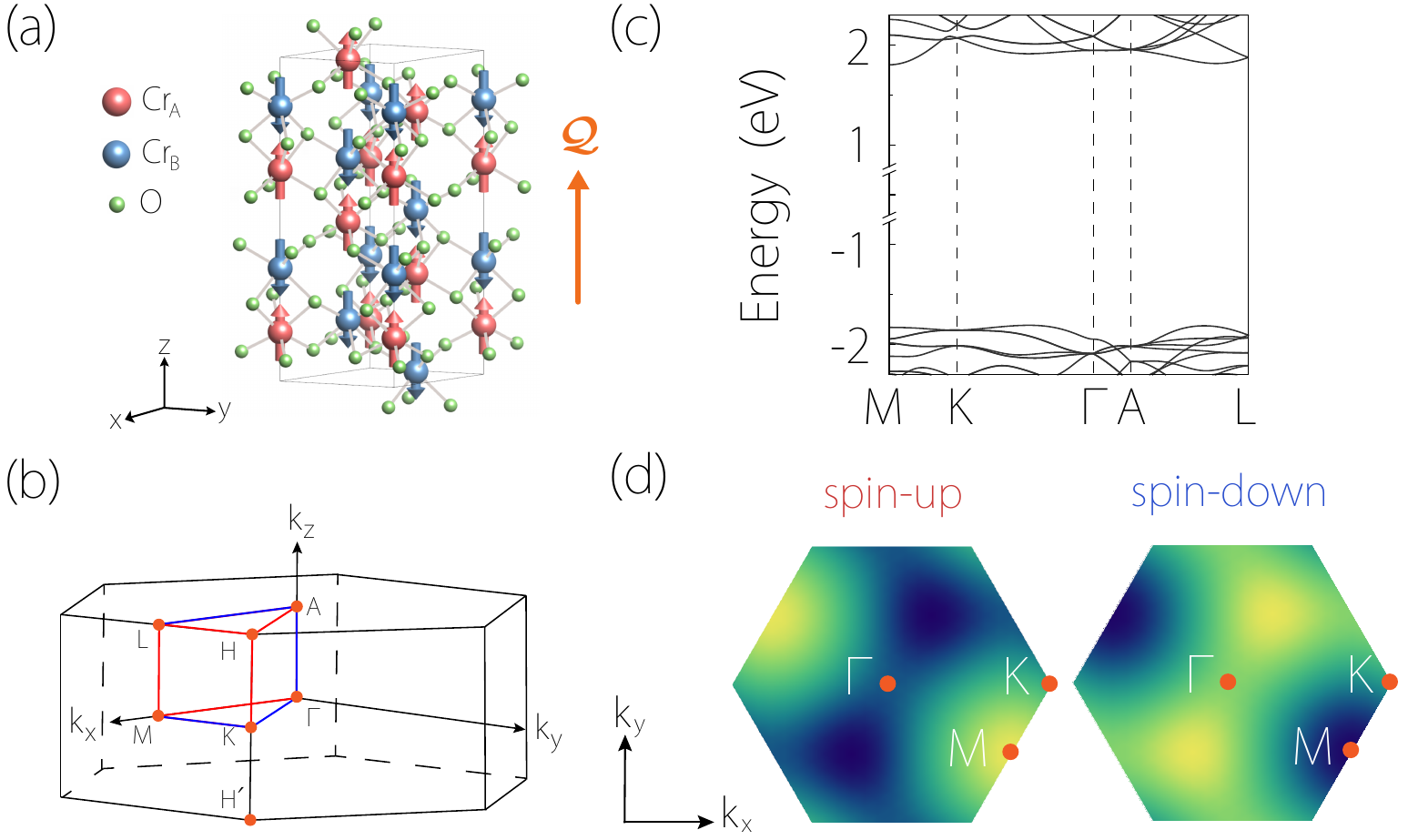}
	\caption{(a) Crystalline  structure of  $\mathrm{Cr_{2}O_{3}}$, where the red and blue spheres represent the $\mathrm{Cr}$ atoms with opposite magnetic moments. It is an $\mathcal{P}T$-AFM type-II AFE. (b) BZ of $\mathrm{Cr_{2}O_{3}}$. 
		(c) Band structure of $\mathrm{Cr_{2}O_{3}}$ without SOC, which are spin-degenerate.
		(d) The $k$ resolved polarization $P_z=\int_{0}^{2\pi}{\cal A}_{z}(k_x,k_y) dk_z$ in the $k_x$-$k_y$ plane. It takes opposite values for the two spin subspaces, leading to AFE.
	}
	\label{figs7}
\end{figure}

The $\mathrm{Cr_{2}O_{3}}$ is  an experimentally synthesized  material \cite{He2024a}. 
$\mathrm{Cr_{2}O_{3}}$ has a trigonal lattice structure with optimized lattice constant $a=b=4.96$ \AA ~and $c=13.59$ \AA~\cite{He2024a}, as shown in Fig. \ref{figs7}(a).  The local moments are mainly on the Cr sites, with an easy axis in the $c$ direction forming a collinear AFM configuration, as illustrated in Fig. \ref{figs7}(a). 
The ground-state magnetic moment of each Cr atom is about  $\pm$2.967 $\mu_{B}$.
In the AFM state, the spin point group is ${}^{\overline{1}}\overline{3} {}^{1}m {}^{\infty m} 1$. According to Table I in the main text, the system is a $\mathcal{P}T$-AFM, which allows AFE order $\bm{\mathcal{Q}}$ in the $z$ direction ($c$ axis).

In Fig. \ref{figs7}(c), we plot the calculated band structure of $\mathrm{Cr_{2}O_{3}}$ in the absence of SOC with \textit{U}=5.5 eV. One observes it is  a magnetic semiconductor with a indirect band gap $\sim 3.61$ eV. The spin degeneracy in the band structure can be clearly seen, manifesting its $\mathcal{P}T$-AFM character.
Using Eqs.~(2) and (3) in the main text, we compute the AFE order parameter $\bm{\mathcal{Q}}$. The result confirms that $\bm{\mathcal{Q}}$ is along $z$ direction, with a value $\sim 14.13$ $\mathrm{\mu C/cm^2}$ [see Fig. \ref{figs7}(d)]. This value is comparable to the ferroelectric polar
ization in perovskite oxides~\cite{FunctionalMetalOxides}.
In addition, we find that  the type-II AFE order of $\mathrm{Cr_{2}O_{3}}$ vanishes above magnetic transition, and is flipped under the reversal of N\'eel vector.
These results reveal  $\mathrm{Cr_{2}O_{3}}$ as a $\mathcal{P}T$-AFM spin-AFE with sizable AFE and strong coupling between AFM and AFE orders.

\subsubsection{$\mathcal{P}T$-AFM material $\mathrm{MgMnO_{3}}$}

\begin{figure}[htbp]
	\centering
	\includegraphics[width=0.7\columnwidth]{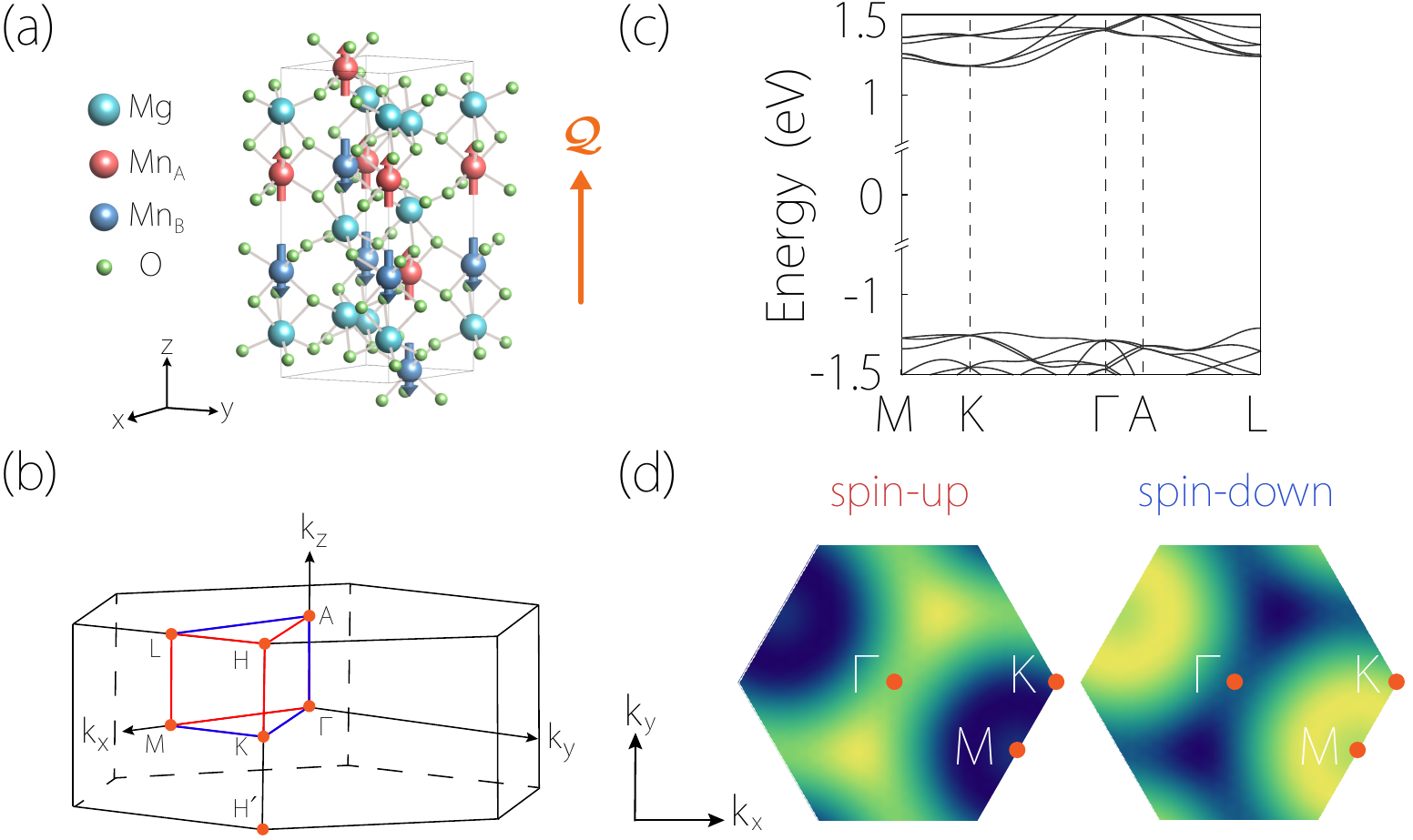}
	\caption{(a) Crystalline  structure of  $\mathrm{MgMnO_{3}}$, where the red and blue spheres represent the $\mathrm{Mn}$ atoms with opposite magnetic moments. It is an $\mathcal{P}T$-AFM type-II AFE.  (b) BZ of $\mathrm{MgMnO_{3}}$. 
		(b) Band structure of $\mathrm{MgMnO_{3}}$ without SOC, which are spin-degenerate.
		(c) The $k$ resolved polarization $P_z=\int_{0}^{2\pi}{\cal A}_{z}(k_x,k_y) d{k_z}$ in the $k_x$-$k_y$ plane. It takes opposite values for the two spin subspaces, leading to AFE.
	}
	\label{figs8}
\end{figure}

The $\mathrm{MgMnO_{3}}$ is  an experimentally synthesized  material~\cite{Haraguchi2019}. 
$\mathrm{MgMnO_{3}}$ has a trigonal lattice structure with optimized lattice constant $a=b=4.98$ \AA~and $c=13.82$ \AA~\cite{Haraguchi2019,Zhang2020a}, as shown in Fig. \ref{figs8}(a).  The local moments are mainly on the Mn sites, with an easy axis in the $c$ direction forming a collinear AFM configuration, as illustrated in Fig. \ref{figs8}(a). 
The ground-state magnetic moment of each Mn atom is about  $\pm$3.094 $\mu_{B}$.
In the AFM state, the spin point group is ${}^{\overline{1}}\overline{3} {}^{\infty m} 1$. According to Table I in the main text, the system is a $\mathcal{P}T$-AFM, which allows AFE order $\bm{\mathcal{Q}}$ in the $z$ direction ($c$ axis).

In Fig. \ref{figs8}(c), we plot the calculated band structure of  $\mathrm{MgMnO_{3}}$ in the absence of SOC with \textit{U}=4 eV. One observes it is  a magnetic semiconductor with a indirect band gap $\sim 2.25$ eV. The spin degeneracy in the band structure can be clearly seen, manifesting its $\mathcal{P}T$-AFM character.
Our calculation  confirms that $\bm{\mathcal{Q}}$ of $\mathrm{MgMnO_{3}}$ is along $z$ direction, with a value $\sim 7.926$ $\mathrm{\mu C/cm^2}$ [see Fig. \ref{figs8}(d)], which is comparable to the ferroelectric polarization in perovskite oxides~\cite{FunctionalMetalOxides}.
Similarly, the type-II AFE order of $\mathrm{MgMnO_{3}}$ vanishes above magnetic transition, and is flipped under the reversal of N\'eel vector.
These results reveal  $\mathrm{Cr_{2}O_{3}}$ as a $\mathcal{P}T$-AFM spin-AFE with sizable AFE and strong coupling between AFM and AFE orders.

\subsubsection{AM material $\mathrm{FeS}$}

\begin{figure}[htbp]
	\centering
	\includegraphics[width=0.6\columnwidth]{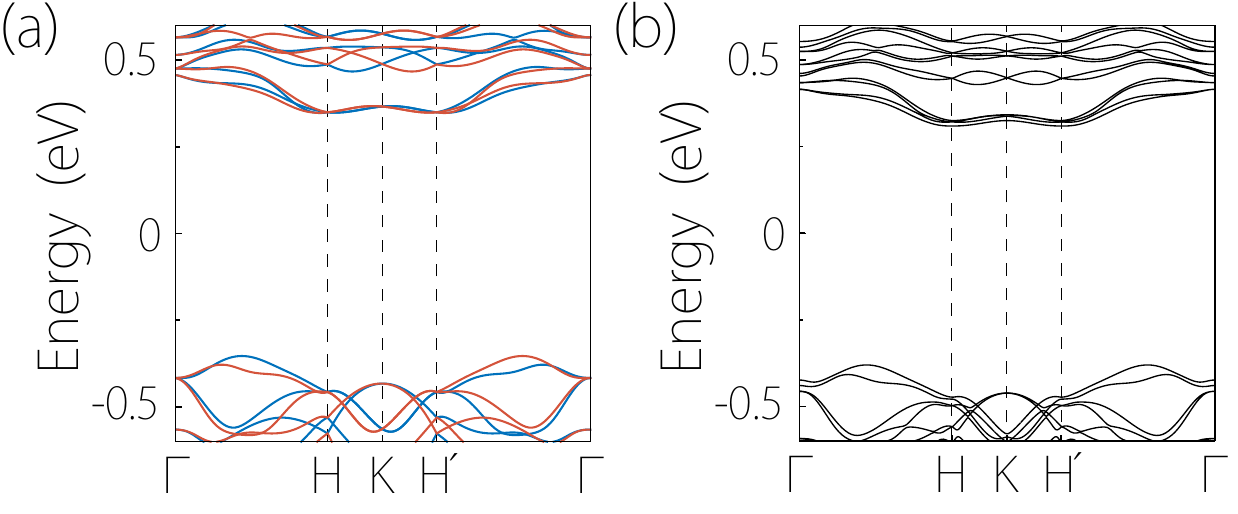}
	  \caption{Band structure of  $\mathrm{FeS}$ (a) without SOC  and (b) with SOC.
}
	\label{figs10}
\end{figure}

The FeS is  an experimentally synthesized  material~\cite{Knusn,Takagi2025}. 
The results of FeS have been presented in the main text. Here, to demonstrate that the effect of SOC on the band structure is negligible, we calculate the band structure of FeS in presence of SOC, as shown in Fig. \ref{figs10}. All calculations employ a Hubbard \textit{U} = 1 eV for Fe atoms and a $15\times15\times6$ \textit{k}-mesh for AFE order parameter determination. The energy-minimized configuration yields a magnetic moment of $\pm$3.11 $\mu_{B}$ for  Fe atoms.

\subsection{2D systems}

\subsubsection{$\mathcal{P}T$-AFM material monolayer $\mathrm{MoICl_{2}}$}

\begin{figure}[htbp]
	\centering
	\includegraphics[width=0.3\columnwidth]{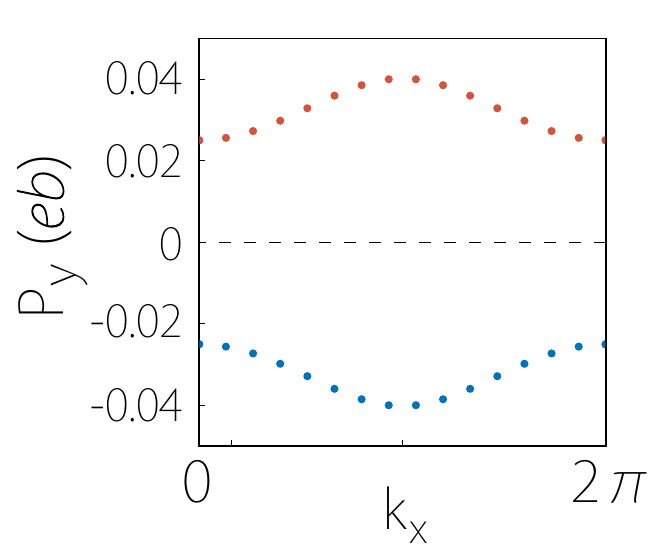}
	\caption{The $k$ resolved polarization $P_y=\int_{0}^{2\pi}{\cal A}_{y}(k_x) dk_y$ for different $k_x$. 
It takes opposite values for the two spin subspaces, leading to AFE. The red and blue dots denote $P_y$ for spin-up and spin-down, respectively.
	}
	\label{figs11}
\end{figure}

The results of monolayer $\mathrm{MoICl_{2}}$ have been presented in the main text.  All calculations employ a Hubbard \textit{U} = 0 eV for Mo atoms and a $15\times9\times1$ \textit{k}-mesh for AFE order parameter determination. The energy-minimized configuration yields a magnetic moment of  $\pm$2.353 $\mu_{B}$ for Mo atoms.
Here, we further plot the $k$ resolved polarization $P_y=\int_{0}^{2\pi}{\cal A}_{y}(k_x) dk_y$ of monolayer $\mathrm{MoICl_{2}}$ in Fig. 	\ref{figs11}, which directly shows the AFE feature of monolayer $\mathrm{MoICl_{2}}$.

\subsubsection{$\mathcal{P}T$-AFM material bilayer $\mathrm{CrI_{3}}$}

\begin{figure}[htbp]
	\centering
	\includegraphics[width=0.8\columnwidth]{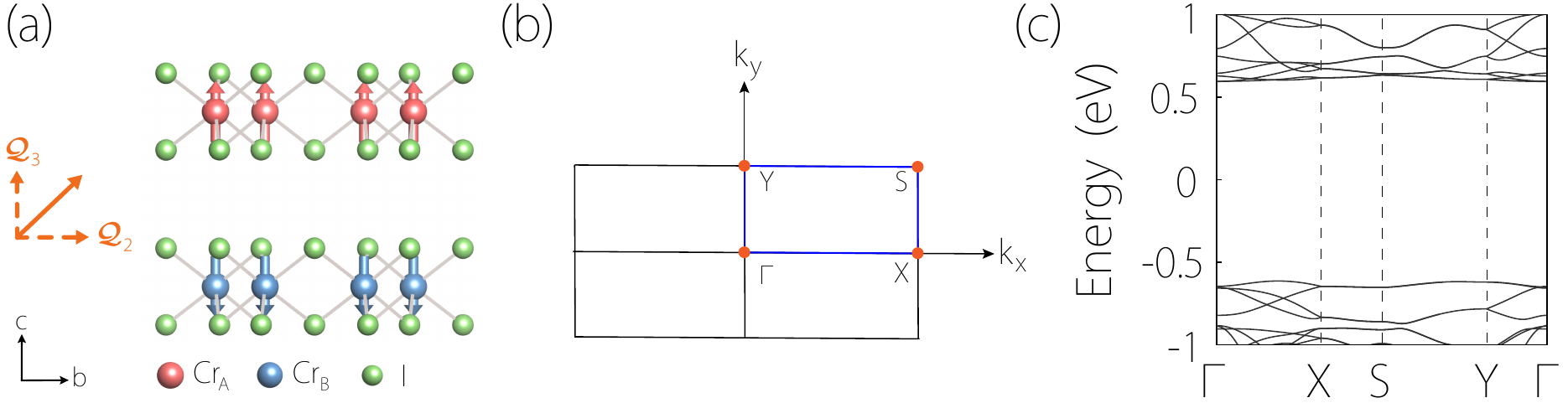}
	\caption{(a) Crystalline  structure of  bilayer $\mathrm{CrI_{3}}$, where the red and blue spheres represent the $\mathrm{Cr}$ atoms with opposite magnetic moments. It is an $\mathcal{P}T$-AFM type-II AFE. (b) BZ of bilayer $\mathrm{CrI_{3}}$.
		(c) Band structure of bilayer $\mathrm{CrI_{3}}$ without SOC, which are spin-degenerate.	}
	\label{figs9-1}
\end{figure}

The bilayer $\mathrm{CrI_{3}}$  is  an experimentally synthesized  material~\cite{Huang}. 
Bilayer $\mathrm{CrI_{3}}$ has an orthorhombic lattice structure with optimized lattice constant $a=7.01$ \AA, and $b=12.1$ \AA ~\cite{Huang,Sivadas2018}, as shown in Fig. \ref{figs9-1}(a).  The local moments are mainly on the Cr sites, with an easy axis in the $c$ direction forming a collinear AFM configuration, as illustrated in Fig. \ref{figs9-1}(a).
The energy-minimized configuration yields a magnetic moment of $\pm$3.042 $\mu_{B}$ for Cr atoms.
In the AFM state, the spin layer-point group is $[{}^{\overline{1}} 2/{}^{1} m{}^{\infty m} 1]_2$. According to Table I in the main text, the system is a $\mathcal{P}T$-AFM, which allows AFE order $\bm{\mathcal{Q}}$ in the $b$-$c$ plane.

\begin{figure}[htbp]
	\centering
	\includegraphics[width=0.6\columnwidth]{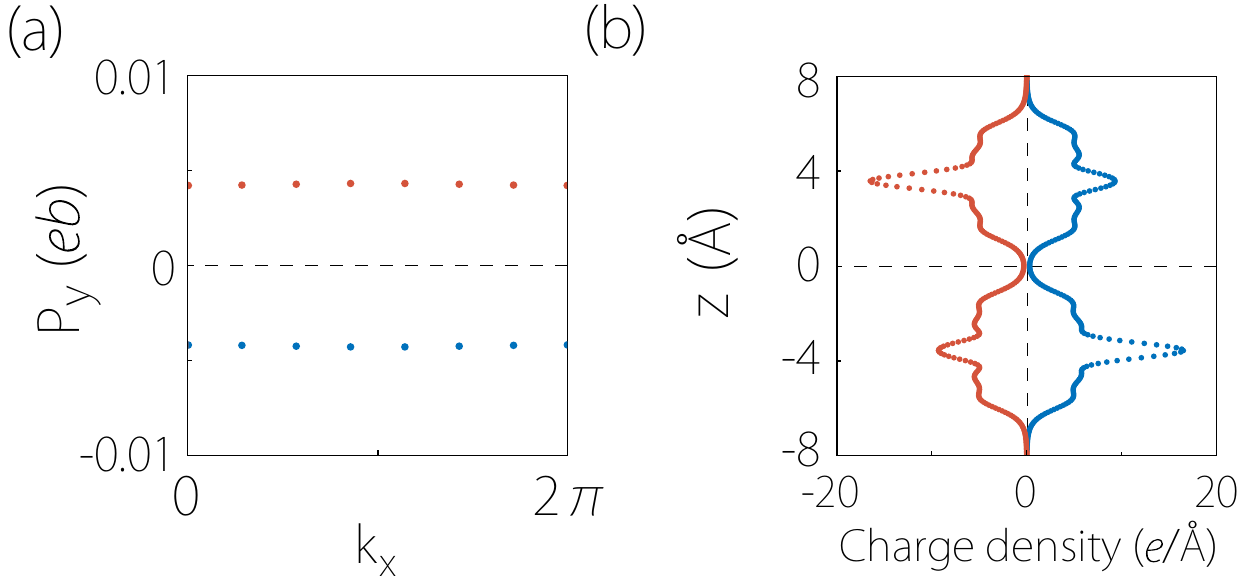}
	\caption{(a) The $k$ resolved polarization $P_y=\int_{0}^{2\pi}{\cal A}_{y}(k_x) dk_y$ for different $k_x$.
		(b) The distribution of the spin-resolved charge density  along $z$ ($c$) direction. Both (a) and (b) show that $\bm{\mathcal{Q}}$ take opposite values for the two spin subspaces, leading to AFE.
The red and blue dots denote the polarization for spin-up and spin-down, respectively.	
}
	\label{figs9-2}
\end{figure}

In Fig. \ref{figs9-1}(c), we plot the calculated the band structure of  bilayer $\mathrm{CrI_{3}}$ in the absence of SOC with \textit{U}=0 eV. One observes it is  a magnetic semiconductor with a band gap $\sim 0.67$ eV. The spin degeneracy in the band structure can be clearly seen, manifesting its $\mathcal{P}T$-AFM character.
Our calculation gives  $\mathcal{Q}_2\sim 1.2$ pC/m and $\mathcal{Q}_3\sim 1615.31$ pC/m, respectively (see Fig. \ref{figs9-2}).
Notice that since  bilayer $\mathrm{CrI_{3}}$   is confined in $z$-direction, $\mathcal{Q}_3$ can be calculated via the distribution of the spin-resolved charge density  along $z$ direction, as shown in Fig. \ref{figs9-2}(b).

\subsection{Verification of the two conditions for the five materials of spin-AFE}  

We use the following table to explicitly show how the five spin-AFE materials satisfy the two conditions in the main text. For condition (i), we present the group $\mathcal G_0^R$, showing $\mathcal G_0^R$ is polar point group; for condition (ii), we list the operation $X$ that reverses the spin-sector polarizations.

Besides, we take FeS as a concrete example and illustrate how the symmetry constrains its AFE order. FeS has spin point group ${}^{\overline{1}} \overline{6}{}^{1} m{}^{\overline{1}} 2{}^{\infty m} 1$. According to our Table I in the maintext, its spin-sector-preserving group $\mathcal{G}_0$ can be generated by $[E||3^{+}_{001}]$ and $[E||m_{100}]$, its $\mathcal{G}_0^R$ group is the polar point group $C_{3v}$, and the spin-switching operation $X$ can be chosen as $[C_2||\overline{6}^{+}_{001}]$. Consider an arbitrary position $(a,b,c)$ expressed in the basis of lattice vectors. Acting on it with the elements of $\mathcal{G}_0^R$ (i.e. $C_{3v}$) generates six positions. Summing over these six positions yields $(0,0,6c)$, showing that for a given spin sector, an electronic polarization along $z$ is symmetry-allowed. Furthermore, one notes that the spin-switching operation $X$ reverses $z$, the polarizations for the two spin sectors are opposite. Therefore, this leads to the type-II spin-AFE with $\mathcal{Q}$ along $z$ direction.

\begin{table}[H]
	\renewcommand{\arraystretch}{1.2}
	\caption{Table showing how the five spin-AFE materials satisfy the two conditions discussed in the main text. }
	\begin{ruledtabular} %
		\begin{tabular}{cccc}
			Materials & $\mathcal G_0^R$ & Condition (i) & Condition (ii) \tabularnewline
			\hline
			$\mathrm{FeS}$ & $C_{3v}$ & $[E||3_{001}^{+}]$ ($\mathcal{Q}_1=0$, $\mathcal{Q}_2=0$, $\mathcal{Q}_3\neq0$) & $X=[C_2||\overline{6}_{001}^{+}]$ ($P_3^+=-P_3^-$) \tabularnewline
Monolayer $ \mathrm{MoICl_2}$ & $C_{2}$ & $[E||2_{010}]$ ($\mathcal{Q}_1=0$, $\mathcal{Q}_2\neq0$, $\mathcal{Q}_3=0$) & $X=[C_2||\mathcal{P}]$ ($P_2^+=-P_2^-$) \tabularnewline

$\mathrm{Cr_{2}O_{3}}$ & $C_{3v}$ & $[E||3_{001}^{+}]$ ($\mathcal{Q}_1=0$, $\mathcal{Q}_2=0$, $\mathcal{Q}_3\neq0$) & $X=[C_2||\mathcal{P}]$ ($P_3^+=-P_3^-$) \tabularnewline

$\mathrm{MgMnO_3}$ & $C_{3}$ & $[E||3_{001}^{+}]$ ($\mathcal{Q}_1=0$, $\mathcal{Q}_2=0$, $\mathcal{Q}_3\neq0$) & $X=[C_2||\mathcal{P}]$ ($P_3^+=-P_3^-$) \tabularnewline

Bilayer $\mathrm{CrI_3}$ & $C_{s}$ & $[E||m_{100}]$ ($\mathcal{Q}_1=0$, $\mathcal{Q}_2\neq0$, $\mathcal{Q}_3\neq0$) & $X=[C_2||\mathcal{P}]$ ($P_2^+=-P_2^-$, $P_3^+=-P_3^-$) \tabularnewline
		\end{tabular}
	\end{ruledtabular}
	\label{tablesadd2}
\end{table}

\subsection{Phase transition from AFE to FE state for $\mathrm{FeS}$}

\begin{figure}[htbp]
	\centering
	\includegraphics[width=0.8\columnwidth]{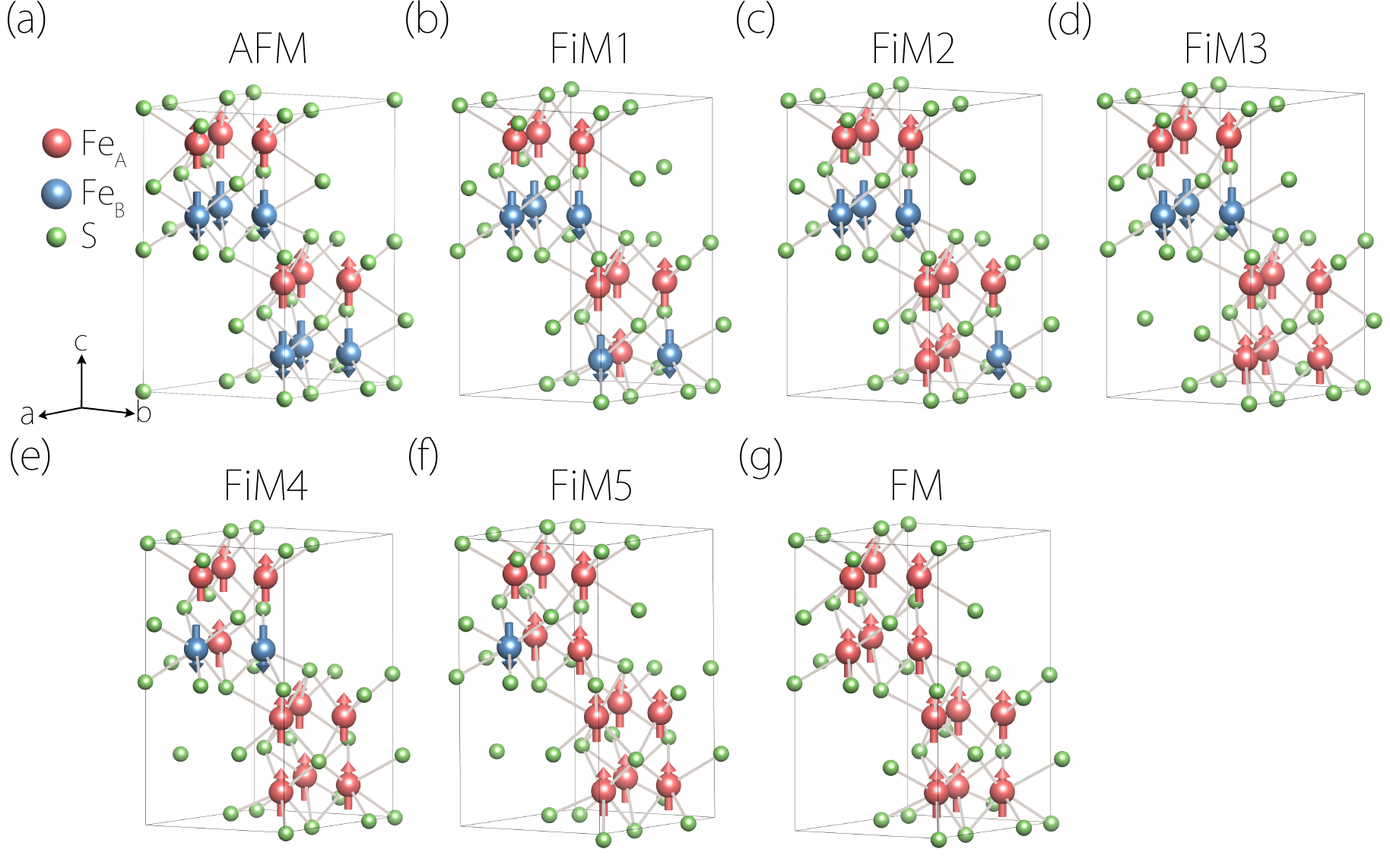}
	  \caption{The different magnetic configurations of $\mathrm{FeS}$.}
	\label{figsadd1}
\end{figure}

\begin{figure}[htbp]
	\centering
	\includegraphics[width=0.7\columnwidth]{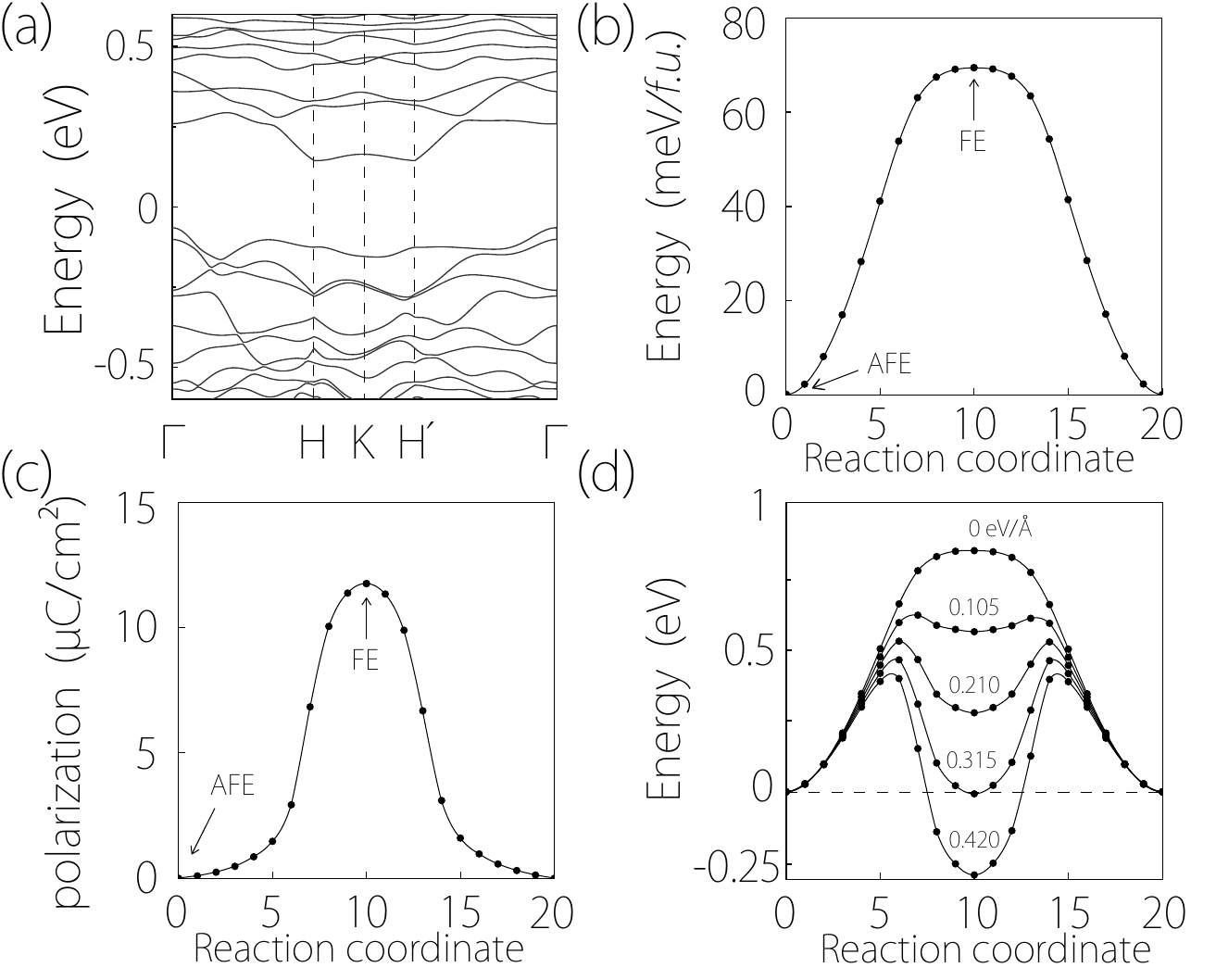}
	  \caption{(a) Band structure of the FiM1 state in presence of SOC. (b) and (c) Energy and polarization variation along the transition path between the AFE state and the FE state in the absence of applied E field. (d) Energy profile of the system under applied E field along the FE polarization direction.}
	\label{figsadd2}
\end{figure}

In the manuscript, we already showed that in type-II AFEs, like $\mathrm{FeS}$, the electric polarization is tied to the magnetic configuration. Hence, in order to determine possible metastable FE states of $\mathrm{FeS}$, we have compared different magnetic configurations [see Fig. \ref{figsadd1} and Table \ref{tablesadd1}]. We find that the ferrimagnetic state FiM1 [Fig. \ref{figsadd1}(b)] is a metastable FE state with a sizable ferroelectric polarization, and its energy is also close to the altermagnetic ground state. The calculated band structure of FiM1 state is shown in Fig. \ref{figsadd2}(a). We further calculate the variation of energy and polarization along the transition path between AFE and FE states (linear interpolation of atomic positions and magnetic moments between the AFE and FE states), as shown in Fig. \ref{figsadd2}(b,c). By applying an electric field along the polarization direction of the FE state, the relative energy of the FE state will be lowered, and the free energy curves of FeS along the transition path under different applied $E$ field are plotted in Fig. \ref{figsadd2}(d). One observes that under a critical value of 0.315 eV/Å, the FE state will become lower than AFE state, indicating that a field-induced transition from AFE state and FE state can happen. Notice that although the critical electric field estimated here is relatively large (0.315 eV/Å), the actual coercive field in experiment generally is much smaller, due to nucleation and domain wall motion. For example, for bulk $\mathrm{CuInP_2S_6}$, the energy barrier and polarization respectively are about 120 meV/f.u. and 2.4 $\mathrm{\mu C/cm^2}$, its reported coercive field is only 77 kV/cm. In contrast, energy barrier for FeS found here is about 69.4 meV/f.u. and the polarization of the FE state of FeS is about 11.751 $\mathrm{\mu C/cm^2}$. Thus, one can expect that the actual coercive field of FeS should be comparable with or smaller than that of bulk  $\mathrm{CuInP_2S_6}$. Therefore, we expect that a double hysteresis loop for FeS can be observed in practice.

\begin{table}[H]
	\renewcommand{\arraystretch}{1.2}
	\caption{The relative free energy, magnitude of electric polarization, and net magnetic moment of FeS under different magnetic configurations (see Fig. \ref{figsadd1}) in the presence of SOC. Their units are meV/f.u., $\mathrm{\mu C/cm^2}$, and $\mu_{B}$, respectively. The states other than AFM and FiM1 are found to be metals, so they do not have a well-defined polarization value. }
	\begin{ruledtabular} %
		\begin{tabular}{cccc}
			Magnetic configuration & Free energy & Polarization & Net magnetic moment  \tabularnewline
			\hline
			AFM & 0 & 0 & 0 \tabularnewline
			FiM1 & 69.4 & 11.75 & 5.8 \tabularnewline
			FiM2 & 117.5 & \textemdash{} & 13.5 \tabularnewline
			FiM3 & 156.1 & \textemdash{} & 19.4 \tabularnewline
			FiM4 & 218.0 & \textemdash{} & 25.0 \tabularnewline
			FiM5 & 263.1 & \textemdash{} & 30.9 \tabularnewline
			FM & 307.7 & \textemdash{} & 35.2 \tabularnewline
		\end{tabular}
	\end{ruledtabular}
	\label{tablesadd1}
\end{table}

\subsection{Energy barrier for switching the AFE order of $\mathrm{FeS}$ and monolayer $\mathrm{MoICl_2}$}

We calculate the energy barrier for switching the AFE order of $\mathrm{FeS}$ and monolayer $\mathrm{MoICl_2}$. The following Fig. \ref{figsadd3} shows the two states of $\mathrm{FeS}$ having opposite AFE order and AFM order. We then calculate the energy barrier along the transition path between the two AFE states (linear interpolation of magnetic moments between the two states). In the calculation, the spin-orbit coupling effect is fully considered, and the result is plotted in Fig. \ref{figsadd4}(a). We find that the energy barrier is relatively small $\sim 0.175$ meV/f.u.. Similarly, we calculate the energy barrier for monolayer $\mathrm{MoICl_2}$, and find that the energy barrier is also small $\sim 0.093$ meV/f.u., as shown in Fig. \ref{figsadd4}(b).

\begin{figure}[H]
	\centering
	\includegraphics[width=0.6\columnwidth]{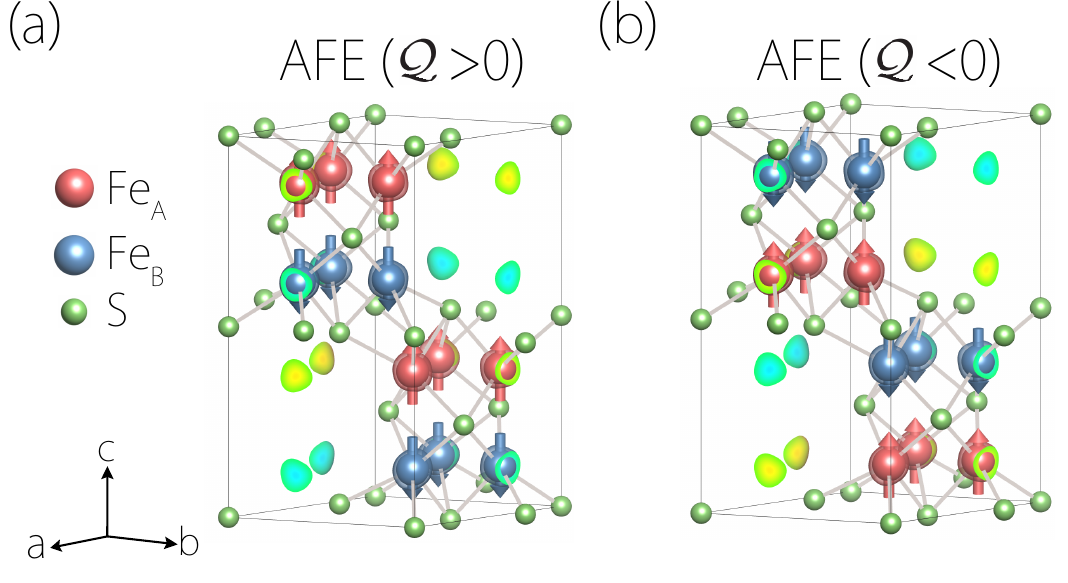}
	\caption{Magnetization-density distributions of $\mathrm{FeS}$ with opposite AFE order.
}
	\label{figsadd3}
\end{figure}

\begin{figure}[H]
	\centering
	\includegraphics[width=0.6\columnwidth]{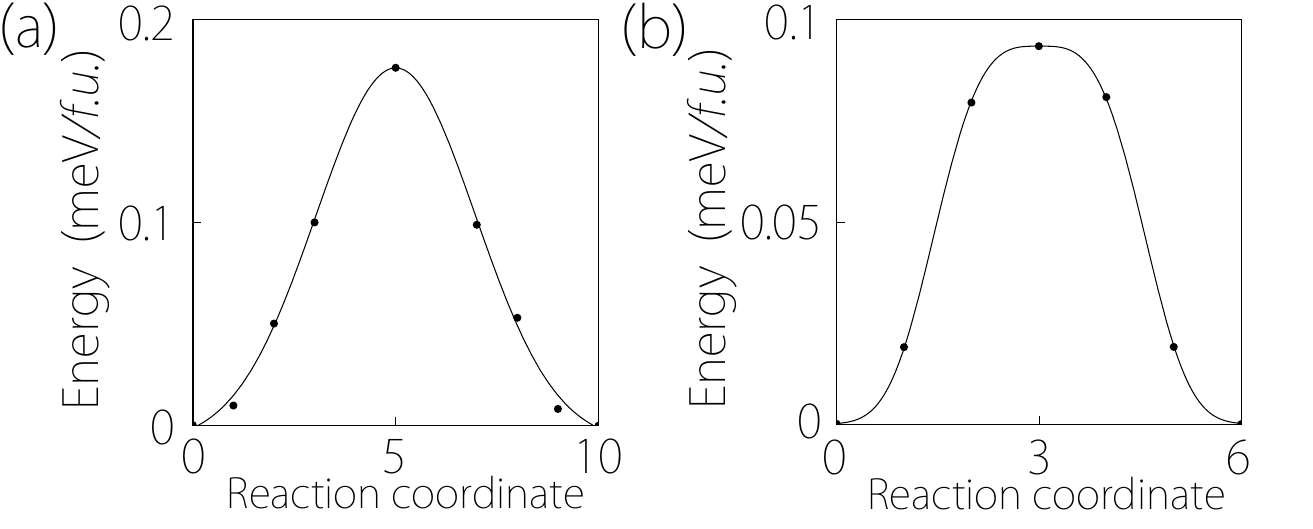}
	\caption{Energy profile along the transition path between the AFE state ($\bm{\mathcal{Q}>0}$) and the AFE state ($\bm{\mathcal{Q}>0}$) state in (a) $\mathrm{FeS}$ and (b) monolayer $\mathrm{MoICl_2}$, respectively.
}
	\label{figsadd4}
\end{figure}

\section{Mirror-AFE in two dimensions}
\subsection{Symmetry condition for mirror-AFE}
As discussed in the main text,  the type-II AFE can be also realized in 2D systems with horizontal mirror symmetry, where $W$ is the horizontal mirror  $m_{001}$, and the
decoupling is into the two subspaces with even and odd mirror eigenvalues.

Similarly,  to achieve a mirror-AFE, the group of the 2D system $\mathcal{G}$ must be decomposed as
\begin{equation}
  \mathcal G=\mathcal G_0+X \mathcal G_0,
\end{equation}
where $\mathcal G_0$ is the set of symmetries preserving each mirror sector, and $X\mathcal G_0$ are those switching the two sectors.
Specifically, we denote the state with  even  (odd) mirror eigenvalues as $|m_z=m\rangle$ ($|m_z=-m\rangle$). Here, $m=1$ for spinless systems and $m=i$ for spinful systems.
Since $X$ connect the two mirror subspaces, it must satisfy the following condition
\begin{eqnarray}\label{P1}
 X |\pm m\rangle \propto |\mp m\rangle.
\end{eqnarray}
This condition indicates that if $X$ is a spatial operator, it must anticommute with $m_{001}$, and if $X$ is an operator containing $T$, it must anticommute (commute) with $m_{001}$ for spinless (spinful) systems, as $T$ connects the two states whose eigenvalues are complex conjugates of each other. 
Besides, the $X$ should  ensures the net polarization vanishes: ${\bm P}={\bm P}^{+}+{\bm P}^{-}=0$.

Based on these two conditions,   it follows that  mirror-AFE cannot exist in  nonmagnetic systems, which have  $T$ symmetry.
For nonmagnetic systems without SOC, i.e. spinless systems,  $T$ symmetry and all point-group operations fail to switch the two mirror sectors.
Moreover, for nonmagnetic systems with SOC, i.e. spinful systems, although  $T$ symmetry  now enables switching the two mirror sectors, it simultaneously enforces same polarization in both sectors, thereby forbidding  the appearance  of mirror-AFE.

Then, we consider the magnetic systems both with and without SOC. For magnetic systems  without SOC, we search through the 94 collinear spin layer-point groups (see Sec. \ref{2DSPG}). Interestingly, due to the decoupling of spin and lattice, there are four spin-group  operators that can be considered as mirror, i.e. 
\begin{eqnarray}\label{P1}
[E|| m_{001}], \ \  [C_2|| m_{001}], \ \  [C_{\|,\infty}|| m_{001}], \ \ [C_2 C_{\|,\infty}|| m_{001}].
\end{eqnarray}
 Here, $C_{\|,\infty}$ is  the  rotation along spin  direction. 
However, we find for all the four mirror operators, there is no spin group that can accommodate mirror-AFE, as  there always exists a symmetry preserving each mirror sector that causes the polarization of the mirror sector to be zero.

\begin{table}[b]
   \renewcommand{\arraystretch}{1.4}
	\caption{ Magnetic layer-point groups  for mirror-AFE, corresponding to the magnetic systems  with SOC. The last column presents the corresponding magnetic layer groups (MLGs).}
	\begin{ruledtabular} %
		\begin{tabular}{cccccccc}
			$\mathcal{G}$ & Generators of $\mathcal{G}_0$  & Symmetry $X$   & Magnetic order & $\bm{\mathcal{Q}}$  &    MLGs     \tabularnewline
			\hline
		 $2^{\prime}/m$  & $m_{001}$ &    $\mathcal{P}T$ &    $\mathcal{P}T$-AFM & $(\mathcal{Q}_1,\mathcal{Q}_2,0)$ &   $6.3.23$    \tabularnewline
\hline
		 & & & & &  $37.7.236$, $47.7.336$   \tabularnewline

        $m^{\prime}mm$ & $m_{001},~2_{010}T$  &  $\mathcal{P}T$   & $\mathcal{P}T$-AFM  & $(0,\mathcal{Q}_2,0)$ &    $40.7.269$, $40.9.271$  \tabularnewline

          & & & & & $44.7.313$, $44.7.313$ &\tabularnewline

		\end{tabular}
	\end{ruledtabular}
	\label{table2}
\end{table}

For magnetic systems  with SOC, we search through the 125 magnetic layer-point groups (see Sec. \ref{MPG}), which describe the point-group symmetry of the 2D systems that has a  finite  thickness along out-of-plane direction,  and  find only 2 groups that can accommodate mirror-AFE. These candidate groups are listed in Table \ref{table2}. 
We observe that  as expected,  these two groups that can host mirror-AFE correspond to magnetic  systems.  Particularly, both  groups are  $\mathcal{P}T$-symmetric antiferromagnetics (AFMs), where  $X$ correspond to $\mathcal{P}T$ symmetry.

These results (Table \ref{table2}) further demonstrate that the type-II AFE materials are intrinsically multiferroic, with both AFE and AFM orderings.

\subsection{Lattice model for mirror-AFE}

\begin{figure}[htbp]
	\centering
	\includegraphics[width=0.8\columnwidth]{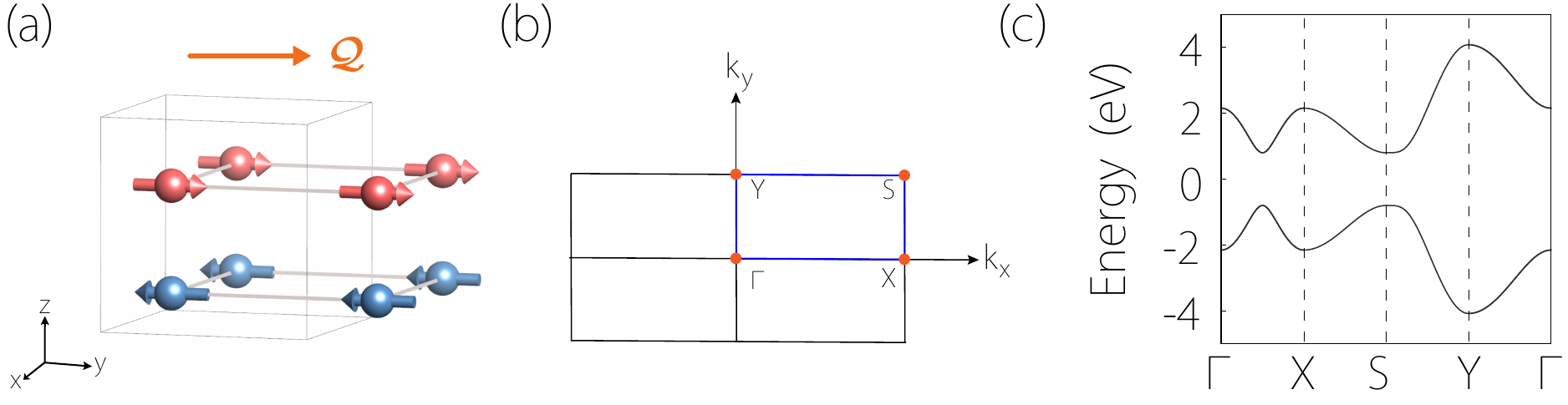}
	\caption{(a) Illustration of the mirror-AFE lattice model (\ref{eq:s2p}). The red and blue arrows denote the magnetic moments on the sites. (b) BZ and (c) band structure  of lattice model (\ref{eq:s2p}). The bands are spin degenerate due to the $\mathcal{P}T$ symmetry. 
Here,  we set $\Delta_{0}=-1$ eV, $t_4 =0.8$ eV, $t_5 =0.8$ eV, $\chi_1=-1$ and $\chi_2 = 2$.}
	\label{figs1-1}
\end{figure}

We construct a simple  lattice model for mirror-AFE with SOC.  We  choose a 2D orthorhombic lattice  belonging to magnetic layer-point group $m^{\prime}mm$  (MLG $37.7.236$) (see Table~\ref{table2}). 
This lattice is  defined on the $x$-$y$ plane, as shown in Fig. \ref{figs1-1}(a). For simplicity, we assume that the lattice constants for this lattice is $a=b=1$.
Each unit of the lattice  contains two  sites  at positions $A=(0,0,z)$ and $B=(0,0,-z)$, where the value of  $z$ is not relevant here.
The two sites each has two spin orbitals, and feature an AFM order, as illustrated in  Fig. \ref{figs1-1}(a).

The lattice Hamiltonian is  constrained by the generators of MLG $37.7.236$, which can be chosen as   $m_{001}$, $\mathcal{P}T$ and $2_{010}T$. 
With the basis states in Fig. \ref{figs1-1}(a), the  symmetry operators are represented as
\begin{eqnarray}\label{eq:s1}
&&m_{001}=-i\tau_{1} \sigma_{3},\ \ \  \mathcal{P}T=i\tau_{1}\sigma_{2}{\cal K},\ \ \   2_{010}T=-\tau_{1}\sigma_{0}{\cal K},
\end{eqnarray}
where $\mathcal{K}$ represents the complex conjugation operator, $\tau$'s and $\sigma$'s  are  Pauli matrices acting on site and spin spaces respectively. $\tau_{0}$ ($\sigma_{0}$) represents the $2 \times 2$ identity matrix.
Following the standard approach \cite{Wieder2016,Yu2019}, the symmetry-allowed lattice Hamiltonian can be established as \cite{zhangMagneticTBPackageTightbinding2022a,zhangMagneticKPPackageQuickly2023}
\begin{eqnarray}\label{eq:s2}
\mathcal{H}(\bm{k})&=&(t_{1} \cos k_{x}+t_{2} \cos k_{y}+ t_{3} \sin k_{x})\tau_{0}\sigma_{0}+t_{4}\sin k_{y}\tau_{3}\sigma_{1}+t_5\tau_{1}\sigma_{0}
\nonumber\\
&&+(\Delta_{0}+t_{6}\cos k_{y}+t_{7} \cos k_{x}+t_{8} \sin k_{x})\tau_{3}\sigma_{2},
\end{eqnarray}
with $t_{1,2,\cdots,8}$ the  hopping parameters and  $\Delta_{0}$ denoting the  exchange term associated with AFM order.
From the exchange term $\Delta_{0}\tau_{3}\sigma_{2}$, one knows that the N\'eel vector is along $y$ direction.
Under a unitary transformation with transformation matrix
\begin{eqnarray}
U=\frac{1}{\sqrt{2}}\begin{pmatrix}
  0 & -1 & 0 & 1\\
  1 & 0 & -1 & 0\\
  0 & 1 & 0 & 1\\
  1 & 0 & 1 & 0
\end{pmatrix},
\end{eqnarray}
the Hamiltonian (\ref{eq:s2}) becomes  a  block-diagonalized matrix, expressed as 
\begin{eqnarray}\label{eq:s2p2}
{\cal H}_{m}(\bm{k})= U \mathcal{H}(\bm{k}) U^{\dagger}&=&(t_{1} \cos k_{x}+t_{2} \cos k_{y}+ t_{3} \sin k_{x})\tau^{\prime}_{0}\sigma^{\prime}_{0}-t_{4}\sin k_{y}\tau^{\prime}_{0}\sigma^{\prime}_{1}+t_5\tau^{\prime}_{3}\sigma^{\prime}_{3}
\nonumber\\
&&+(\Delta_{0}+t_6\cos k_{y}+t_{7} \cos k_{x}+t_{8} \sin k_{x})\tau^{\prime}_{0}\sigma^{\prime}_{2},
\end{eqnarray}
where $\tau^{\prime}$'s and $\sigma^{\prime}$'s  are still Pauli matrices, but no longer act on site and spin spaces.
Notice that here $\tau^{\prime}$'s act on  mirror space. For simplification, we set $t_1=t_2=t_3=t_8=0$ , and the Hamiltonian (\ref{eq:s2p2}) reduces to a simple form
\begin{eqnarray}\label{eq:s2p1}
{\cal H}_{m}(\bm{k})&=&(\Delta_{0}+t_6\cos k_{y}+t_{7} \cos k_{x})\tau^{\prime}_{0}\sigma^{\prime}_{2}-t_{4}\sin k_{y}\tau^{\prime}_{0}\sigma^{\prime}_{1}+t_5\tau^{\prime}_{3}\sigma^{\prime}_{3}.
\end{eqnarray}
Since the first term in (\ref{eq:s2p1}) is odd under $T$ symmetry,  we can rewrite (\ref{eq:s2p1}) to highlight  the role of AFM order. The modified Hamiltonian reads
\begin{eqnarray}\label{eq:s2p}
	{\cal H}_{m}(\bm{k})&=&\Delta_{0}(1+\chi_{1} \cos k_{y}+ {\chi}_2 \cos k_{x})\tau^{\prime}_{0}\sigma^{\prime}_{2}-t_{4}\sin k_{y}\tau^{\prime}_{0}\sigma^{\prime}_{1}+t_5\tau^{\prime}_{3}\sigma^{\prime}_{3},
\end{eqnarray}
where $\chi_{1(2)}$ are dimensionless constants. The band structure of model (\ref{eq:s2p}) is shown in Fig.~\ref{figs1-1}(b), showing the characteristic spin degeneracy  for $\mathcal{P}T$-AFM.

\begin{figure}[htbp]
	\centering
	\includegraphics[width=0.6\columnwidth]{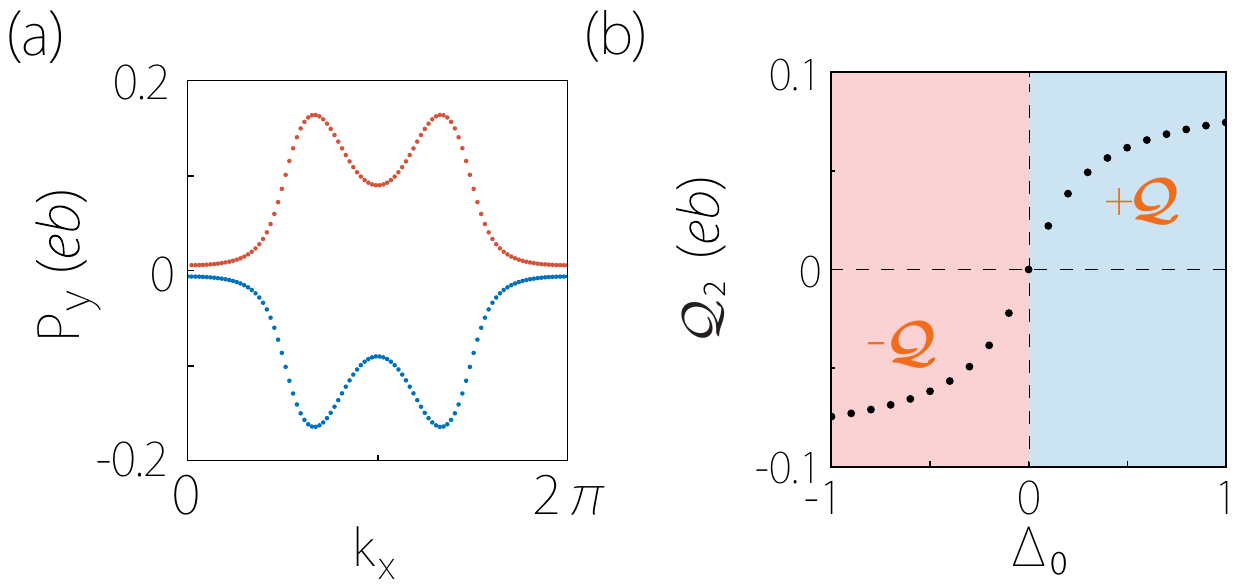}
	\caption{(a) The $k$ resolved polarization $P_y=\int_{0}^{2\pi}{\cal A}_{y}(k_x) dk_y$ for different $k_x$, obtained from  model (\ref{eq:s2p}). It takes opposite values for the two mirror subspaces, leading to mirror-AFE.  The red (blue) dots denotes the results for $m=i$  ($m=-i$)  subspace.
(b) The AFE order $\mathcal{Q}_2$ vs the AFM order $\Delta_0$. Here,  we set $\Delta_{0}=-1$ eV, $t_4 =0.8$ eV, $t_5 =0.8$ eV, $\chi_1=-1$ and $\chi_2 = 2$ in (a-b) and $\Delta_{0}= 1$ eV in (b).}
	\label{figs1-2}
\end{figure}

After computing the polarization of each mirror subspace, we find that the mirror-AFE order  $\bm{\mathcal{Q}}$ is along the $y$ axis, with ${\mathcal{Q}}_2=0.075\ eb$ ($b$ is  the lattice constant) [see Fig. \ref{figs1-2}(a)].
Similarly,  nonzero $\Delta_0$ is also essential for finite mirror-AFE order [see Fig. \ref{figs1-2}(b)], demonstrating that mirror-AFE results from the AFM order. By  reversing the N\'eel  vector, the mirror-AFE order $\bm{\mathcal{Q}}$ is also reversed [see Fig. \ref{figs1-2}(b)], showing  that mirror-AFE materials are intrinsically multiferroic with strong magnetoelectric coupling.

\begin{figure}[htb]
	\centering
	\includegraphics[width=0.63\columnwidth]{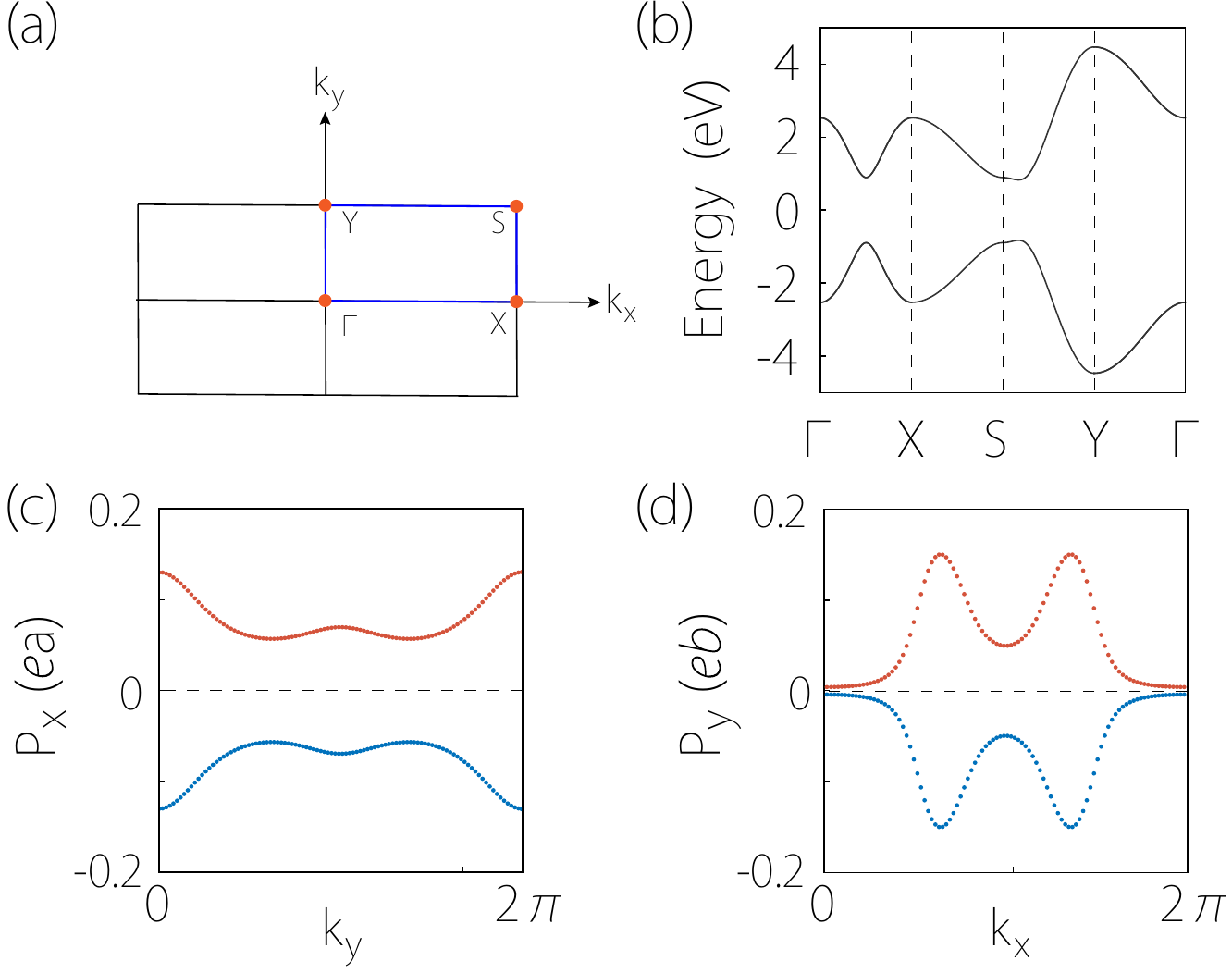}
	   \caption{(a) BZ and (b)  band structure of the lattice model (\ref{eq:s4}), which are spin degenerate. (c) The $k$ resolved polarization $P_x=\int_{0}^{2\pi}{\cal A}_{x}(k_y) dk_x$ for different $k_y$. (d) $P_y=\int_{0}^{2\pi}{\cal A}_{y}(k_x) dk_y$ for different $k_x$. Both $P_x$ and $P_y$ take opposite values for the two mirror subspaces, leading to mirror-AFE. The red (blue) dots denotes the results for $m=i$  ($m=-i$)  subspace. 
	   Here,  we set $\Delta_{0}=-1$ eV, $t_4 =0.8$ eV, $t_5 =0.8$ eV, $\chi_1=-1$, $\chi_2 = 2$ and $t^{\prime}=0.4$ eV.}
	\label{Fig.S2}
\end{figure}

Besides, if we  break the $2_{010}T$ symmetry but preserve  $m_{001}$ and $\mathcal{P}T$,  the magnetic layer-point group of the lattice model becomes $2'/m$, and then the model will  have finite AFE polarization along  $x$ direction. To directly show it, we add a  $2_{010}T$-breaking  term  in the Hamiltonian (\ref{eq:s2p}), and the new Hamiltonian is  
\begin{eqnarray}\label{eq:s4}
{\cal H}_m'(\bm{k})&=&{\cal H}_{m}(\bm{k})+t^{\prime} \sin k_{x}\tau^{\prime}_{0}\sigma^{\prime}_{1}- t^{\prime} \cos k_{x}\tau^{\prime}_{0}\sigma^{\prime}_{2}.
\end{eqnarray} 
The calculated band structure and $k$ resolved polarization (Wilson loops) for different mirror subsystems  are shown in Fig. \ref{Fig.S2}.
By integrating the Wilson loops, we obtain the AFE polarization, which is $\bm{ \mathcal{Q}}=(0.078 e a, 0.058 e b,0)$.

\subsection{Material candidates of mirror-AFE}

Guided by  Table \ref{table2}, we find several material candidates for mirror-AFE, including Ruddlesden-Popper related iron oxyhalides: monolayer $\mathrm{Sr_{3}X_{2}Cl_{2}O_{5}}$ ($\mathrm{X=Fe,Ru,Os)}$. The  3D bulk material $\mathrm{Sr_{3}Fe_{2}Cl_{2}O_{5}}$ has already been experimentally synthesized and proven to be AFM at 300 K \cite{Knee2004,1984,Ackerman1991}.

\begin{figure}[H]
	\centering
	\includegraphics[width=0.6\columnwidth]{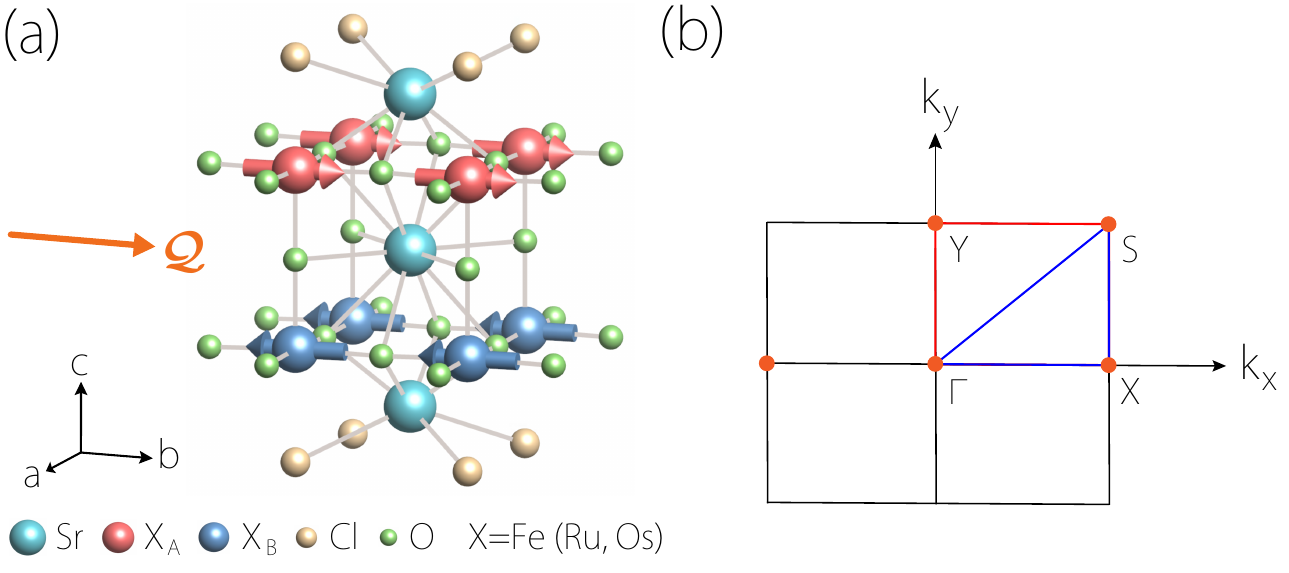}
	\caption{(a)  Crystalline structure and (b) BZ of monolayer $\mathrm{Sr_{3}X_{2}Cl_{2}O_{5}}$ $\mathrm{X=(Fe,Ru,Os)}$.  The red and blue arrows denote the magnetic moments on the X sites.
	}
	\label{figs4-1}
\end{figure}

\begin{figure}[H]
	\centering
	\includegraphics[width=0.8\columnwidth]{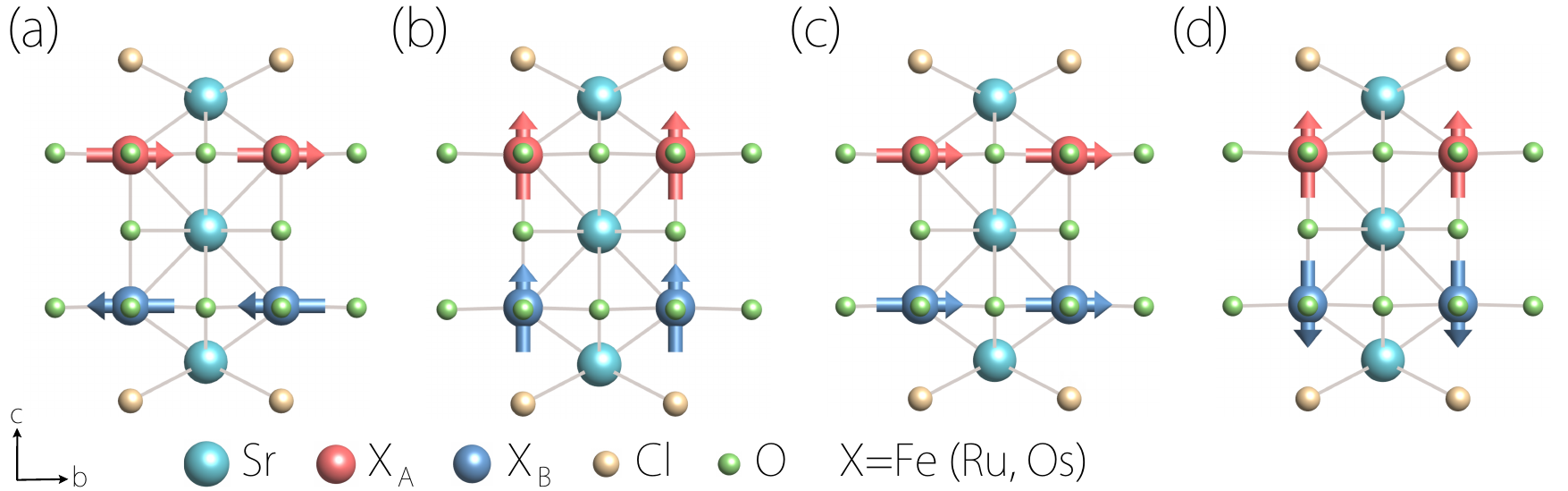}
	  \caption{Illustrations of typical magnetic configurations. The ground state has 010-AFM  ordering. (a) 010-AFM, (b) 001-FM, (c) 010-FM, (d) 001-AFM.
}
	\label{Fig.S12}
\end{figure}

Monolayer $\mathrm{Sr_{3}X_{2}Cl_{2}O_{5}}$ $\mathrm{(X=Fe,Ru,Os)}$ has a square lattice structure, consisting of seven atomic layers in the sequence of $\mathrm{Cl\textendash Sr\textendash X(O)\textendash Sr(O)\textendash X(O)\textendash Sr\textendash Cl}$, as shown in Fig. \ref{figs4-1}(a). 
Transition metal ions with partially filled d shells are typical sources for magnetism. In the monolayer $\mathrm{Sr_{3}X_{2}Cl_{2}O_{5}}$ ($\mathrm{X=Fe,Ru,Os)}$, the
magnetism is mainly from X ions. To determine the magnetic order, we   compare the  energies of different magnetic configurations, which  010-AFM, 001-ferromagnetic (FM), 010-FM, 001-AFM (see Fig. \ref{Fig.S12}). Our calculations show that  the 010-AFM configuration as the lowest-energy state across all \textit{U} values, as shown in Tables \ref{tables4}-\ref{tables2}.

In the 010-AFM state, the magnetic layer-point group of monolayer $\mathrm{Sr_{3}X_{2}Cl_{2}O_{5}}$  is $m^{\prime}mm$ (magnetic layer group 37.7.236). According to Table \ref{table2}, the systems are   $\mathcal{P}T$-AFM, which allows AFE order $\bm{\mathcal{Q}}$ in the $y$ direction ($b$ axis).

\begin{table}[H]
   \renewcommand{\arraystretch}{1.2}
	\caption{The relative energies (in unit of eV/$\mathrm{Fe}$ atom) of monolayer $\mathrm{Sr_{3}Fe_{2}Cl_{2}O_{5}}$ for 010-AFM, 001-FM, 010-FM and 001-AFM. The  ground-state energy for 010-AFM is chosen as a reference.}
	\begin{ruledtabular} %
		\begin{tabular}{cccccc}
			 $U (eV)$ & 010-AFM & 001-FM & 010-FM & 001-AFM \tabularnewline
			\hline

 4.3 & 0 & 0.109031 & 0.108855 & 0.000335\tabularnewline
 4.8 & 0 & 0.100806 & 0.100656 & 0.000306\tabularnewline
 5.3 & 0 & 0.092925 & 0.092797 & 0.000273\tabularnewline

		\end{tabular}
	\end{ruledtabular}
	\label{tables4}
\end{table}

\begin{table}[H]
	\renewcommand{\arraystretch}{1.2}
	\caption{The relative energies (in unit of eV/$\mathrm{Ru}$ atom) of monolayer $\mathrm{Sr_{3}Ru_{2}Cl_{2}O_{5}}$ for 010-AFM, 001-FM, 010-FM and 001-AFM. The  ground-state energy for 010-AFM is chosen as a reference.}
	\begin{ruledtabular} %
		\begin{tabular}{cccccc}
			$U (eV)$ & 010-AFM & 001-FM & 010-FM & 001-AFM \tabularnewline
			\hline
			5.3 & 0 & 0.112986 & 0.196976 & 0.374659 \tabularnewline
			5.5 & 0 & 0.144614 & 0.478063 & 0.383122 \tabularnewline
			6.0 & 0 & 0.155520 & 0.446599 & 0.350848 \tabularnewline

		\end{tabular}
	\end{ruledtabular}
	\label{tables6}
\end{table}

\begin{table}[H]
	\renewcommand{\arraystretch}{1.2}
	\caption{The relative energies (in unit of eV/$\mathrm{Os}$ atom) of monolayer $\mathrm{Sr_{3}Os_{2}Cl_{2}O_{5}}$ for 010-AFM, 001-FM, 010-FM and 001-AFM. The  ground-state energy for 010-AFM is chosen as a reference. }
	\begin{ruledtabular} %
		\begin{tabular}{cccccc}
			$U (eV)$ & 010-AFM & 001-FM & 010-FM & 001-AFM \tabularnewline
			\hline
			
			3.2 & 0 & 0.127072 & 0.126574 & 0.002055\tabularnewline
			
			3.5 & 0 & 0.182906 & 0.223384 & 0.201736 \tabularnewline
			
			3.8 & 0 & 0.240260 & 0.256092 & 0.282765 \tabularnewline
			
		\end{tabular}
	\end{ruledtabular}
	\label{tables2}
\end{table}

\begin{figure}[H]
	\centering
	\includegraphics[width=0.5\columnwidth]{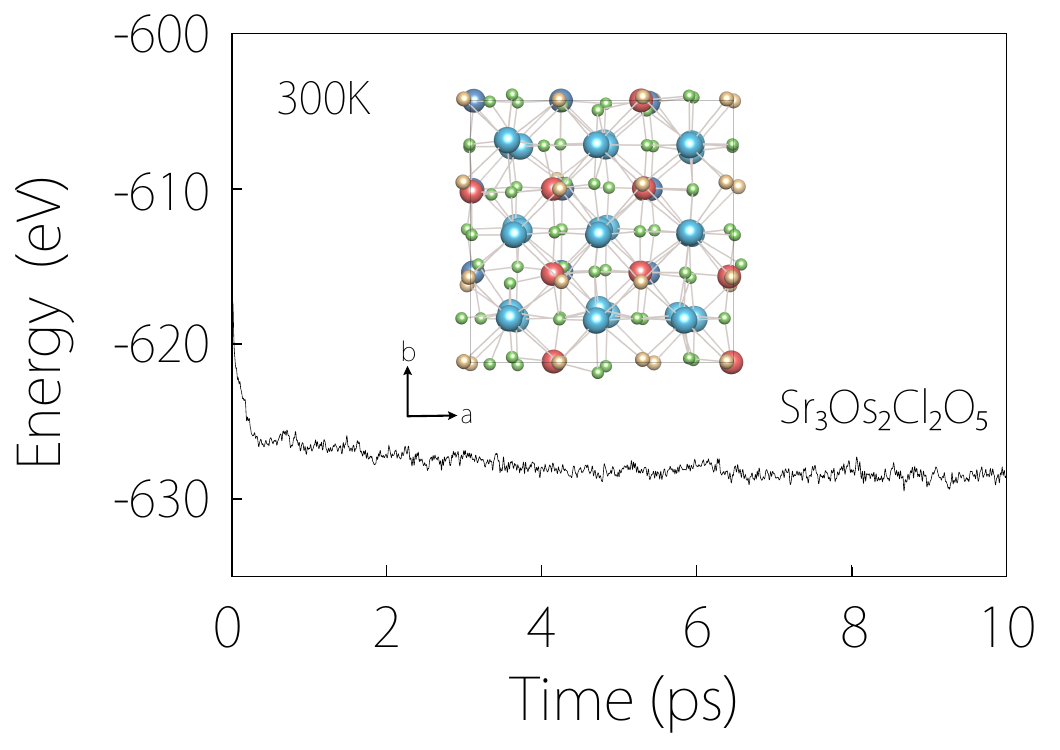}
	\caption{The  total energies  with AIMD simulation by using $3\times3\times1$ supercell at 300 K, and inset shows the snapshots of the crystal structures after 10 ps.
	}
	\label{Fig.S11}
\end{figure}

Next, we  use monolayer $\mathrm{Sr_{3}Os_{2}Cl_{2}O_{5}}$ as an example to check the thermal stability of monolayer $\mathrm{Sr_{3}X_{2}Cl_{2}O_{5}}$. 
The thermal stability of monolayer $\mathrm{Sr_{3}Os_{2}Cl_{2}O_{5}}$ is systematically examined through \textit{ab initio} molecular dynamics (AIMD) simulations conducted at 300 K using a $3\times3\times1$ supercell. Structural snapshots captured during the 10 ps simulation (see Fig. \ref{Fig.S11}) reveal thermally induced lattice distortions with variable amplitudes, yet exhibit no observable bond rupture or structural fragmentation. This quantitative analysis conclusively demonstrates that  the monolayer $\mathrm{Sr_{3}Os_{2}Cl_{2}O_{5}}$  has a robust structure  and thermodynamic resilience under ambient thermal agitation.

\subsubsection{Monolayer $\mathrm{Sr_{3}Os_{2}Cl_{2}O_{5}}$}

Monolayer $\mathrm{Sr_{3}Os_{2}Cl_{2}O_{5}}$ has a square lattice structure with optimized lattice constant $a=b=4.11$ \AA.  The local moments are mainly on the Os sites, 
with a magnitude of $\pm$0.59 $\mu_{B}$ [see Fig. \ref{figs4-1}(a)]. 

\begin{figure}[htbp]
	\centering
	\includegraphics[width=0.7\columnwidth]{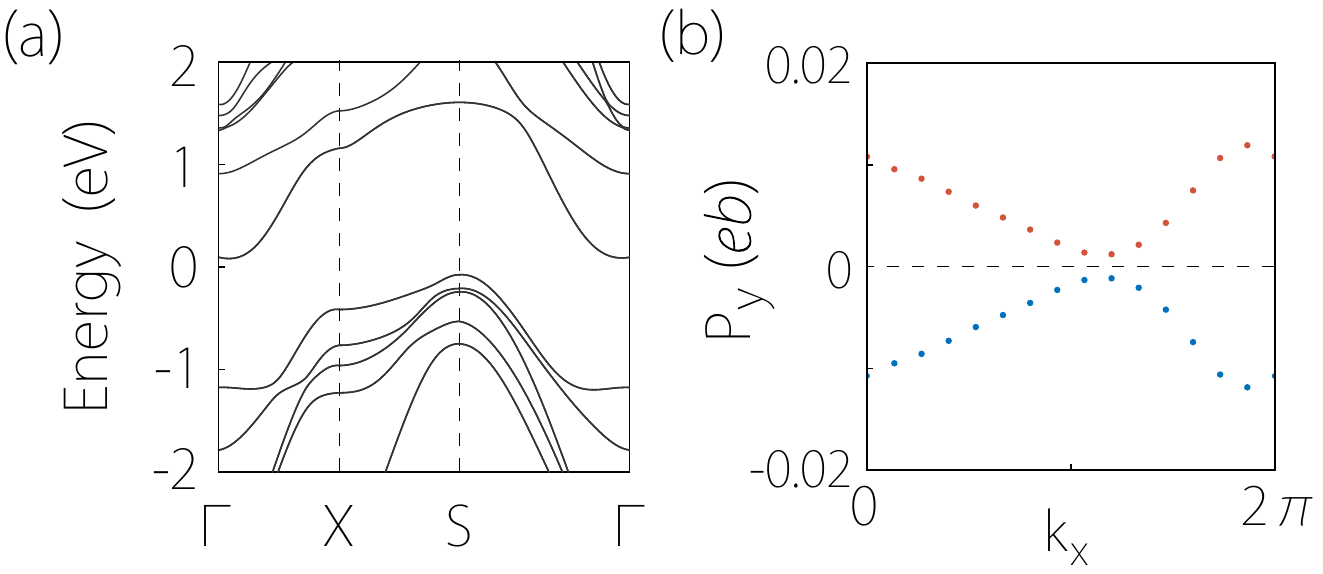}
	\caption{(a) Band structure of monolayer $\mathrm{Sr_{3}Os_{2}Cl_{2}O_{5}}$ with SOC for \textit{U}=3.2 eV, which are spin degenerate. (c) The $k$ resolved polarization $P_y=\int_{0}^{2\pi}{\cal A}_{y}(k_x) dk_y$ for different $k_x$. It takes opposite values for the two mirror subspaces, leading to AFE. The red (blue) dots denotes the results for $m=i$  ($m=-i$)  subspace.
	}
	\label{figs4-2}
\end{figure}

In Fig. \ref{figs4-2}(a), we plot the  band structure of monolayer $\mathrm{Sr_{3}Os_{2}Cl_{2}O_{5}}$ in the presence of SOC with \textit{U}=3.2 eV. One observes it is  a magnetic semiconductor with a indirect band gap $\sim 0.15$ eV. The spin degeneracy in the band structure can be clearly seen, manifesting its $\mathcal{P}T$-AFM character.
Our calculations confirm that $\bm{\mathcal{Q}}$ is along $y$ direction, with a value $\sim 2.39$ pC/m [see Fig. \ref{figs4-2}(b)]. As a comparison, the ferroelectric $\mathrm{WTe_{2}}$ with polarization of  $0.16$ pC/m is detected in experiments \cite{Fei2018}.
In addition,  we find that the type-II AFE order of monolayer $\mathrm{Sr_{3}Os_{2}Cl_{2}O_{5}}$ vanishes above magnetic transition, and is flipped under the reversal of N\'eel vector.
These results reveal  monolayer $\mathrm{Sr_{3}Os_{2}Cl_{2}O_{5}}$ as an $\mathcal{P}T$-AFM mirror-AFE with sizable AFE and strong coupling between AFM and AFE orders.

\subsubsection{Monolayer $\mathrm{Sr_{3}Fe_{2}Cl_{2}O_{5}}$}

Monolayer $\mathrm{Sr_{3}Fe_{2}Cl_{2}O_{5}}$ has a square lattice structure with optimized lattice constant $a=b=3.97$ \AA.  
For  monolayer $\mathrm{Sr_{3}Fe_{2}Cl_{2}O_{5}}$, the magnetic moments are mainly on the Fe atoms with a magnitude of  $\pm$4.325 $\mu_{B}$ for  $U=5.3$ eV.
The band structure of monolayer $\mathrm{Sr_{3}Fe_{2}Cl_{2}O_{5}}$ with SOC  for $U=5.3$ eV is shown in  Fig. \ref{figs6}(a), and the mirror-AFE order parameter $\mathcal{Q}_2$ is calculated as $0.046$ pC/m.

\subsubsection{Monolayer $\mathrm{Sr_{3}Ru_{2}Cl_{2}O_{5}}$}

\begin{figure}[htbp]
	\centering
	\includegraphics[width=0.7\columnwidth]{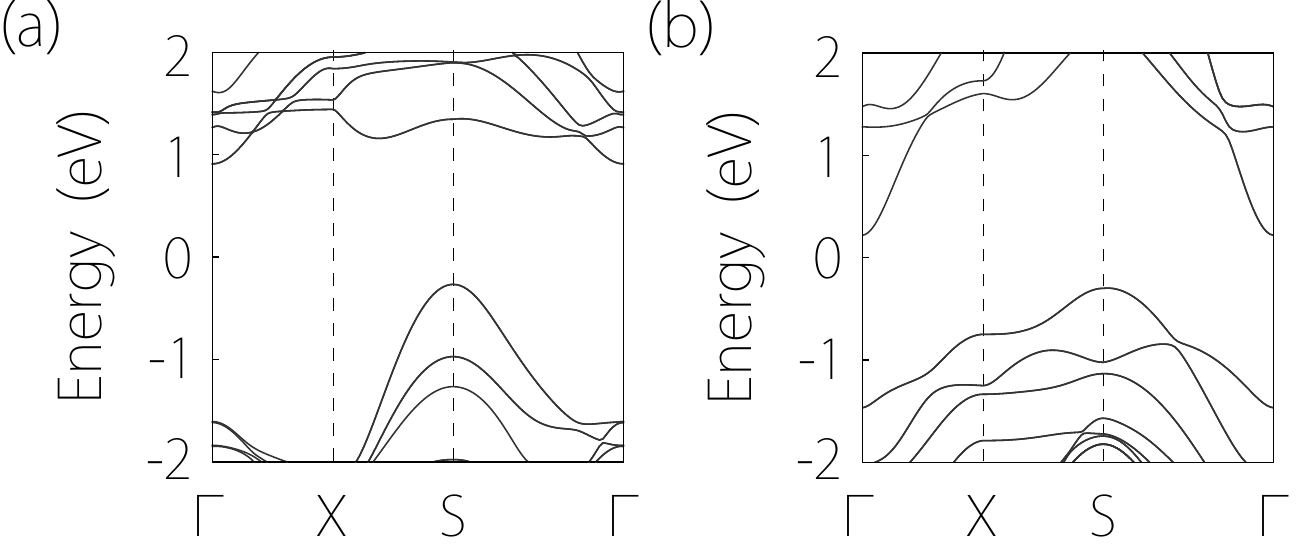}
	  \caption{
Band structure of (a) monolayer $\mathrm{Sr_{3}Fe_{2}Cl_{2}O_{5}}$ and (b)  monolayer $\mathrm{Sr_{3}Ru_{2}Cl_{2}O_{5}}$ with SOC, which are spin degenerate.  Here, we set \textit{U}=5.3 eV for both materials.
}
	\label{figs6}
\end{figure}

The lattice constant of monolayer $\mathrm{Sr_{3}Ru_{2}Cl_{2}O_{5}}$ is $a=b=4.09$ \AA.  
The magnetic moment of this material are mainly on the Ru atoms with a magnitude of  $\pm$2.416 $\mu_{B}$ for  $U=5.3$ eV.
The band structure of monolayer $\mathrm{Sr_{3}Ru_{2}Cl_{2}O_{5}}$ with SOC  for $U=5.3$ eV is shown in  Fig. \ref{figs6}(b), and the mirror-AFE order parameter $\mathcal{Q}_2$ is calculated as $0.687$ pC/m.

\section{Operators of 2D and 3D collinear spin-point groups}
\subsection{Operators  of 94 2D collinear spin layer-point groups}\label{2DSPG}

\begin{table}[H]
	\renewcommand{\arraystretch}{1.6}
	\caption{63 collinear spin layer-point group $\mathcal G$: $\mathcal G=(\mathcal G_0+[C_2T||E]\mathcal G_0)\ltimes SO(2)$ for AFM.  Only nontrivial operators in $\mathcal G_0$  are listed. The generators of $\mathcal G_0$ are marked in blue. The groups allowing for 2D spin-AFE are highlighted by red.  The subscript $i$ in the second column ($[...]_i$) indicates the $i$-th collinear spin layer-point group obtained from a same 3D collinear spin point group. 
}
	\begin{ruledtabular} %
\begin{tabular}{lll}
		No. &	Notation & Operators  in $\mathcal G_0$ [$\mathcal G=(\mathcal G_0+[C_2T||E]\mathcal G_0)\ltimes SO(2)$]   \tabularnewline
\hline
1&  {\color{red}${}^{\overline{1}} \overline{1}{}^{\infty m} 1$}    &   {\color{blue} $[C_{2}||\mathcal{P}]$},~$[E||E]$ 
\tabularnewline

2& $[{}^{\overline{1}} 2{}^{\infty m} 1]_1$    &   {\color{blue} $[C_{2}||2_{001}]$},~$[E||E]$ 
\tabularnewline

3& $[{}^{\overline{1}} 2{}^{\infty m} 1]_2$    &   {\color{blue} $[C_{2}||2_{100}]$},~$[E||E]$ 
\tabularnewline

4& $[{}^{\overline{1}} m{}^{\infty m} 1]_1$    &   {\color{blue} $[C_{2}||m_{001}]$},~$[E||E]$ 
\tabularnewline

5& $[{}^{\overline{1}} m{}^{\infty m} 1]_2$    &   {\color{blue} $[C_{2}||m_{100}]$},~$[E||E]$ 
\tabularnewline

6& {\color{red} $[{}^{1} 2/{}^{\overline{1}} m{}^{\infty m} 1]_1$ }   &   {\color{blue} $[E||2_{001}],~[C_{2}||\mathcal{P}]$},~$[E||E],~[C_{2}||m_{001}]$ 
\tabularnewline

7& {\color{red} $[{}^{1} 2/{}^{\overline{1}} m{}^{\infty m} 1]_2$ }   &   {\color{blue} $[E||2_{100}],~[C_{2}||\mathcal{P}]$},~$[E||E],~[C_{2}||m_{100}]$ 
\tabularnewline
\end{tabular}
	\end{ruledtabular}
	\label{table2DCSPG1}
\end{table}

\begin{table}[H]
	\renewcommand{\arraystretch}{1.6}

	\begin{ruledtabular} %
		\begin{tabular}{lll}
				No. &	Notation & Operators  in $\mathcal G_0$ [$\mathcal G=(\mathcal G_0+[C_2T||E]\mathcal G_0)\ltimes SO(2)$]  \tabularnewline
\hline

32& ${}^{\overline{1}} 4{}^{1} m{}^{\overline{1}} m{}^{\infty m} 1$    &   {\color{blue} $[C_{2}||4^{+}_{001}],~[\mathcal{C}_2||m_{100}],~[E||m_{\overline{1}10}]$},~$[E||E],~[E||2_{001}],~[C_{2}||4_{001}^{-}],~[C_{2}||m_{010}],~[E||m_{110}]$ 
\tabularnewline

33& ${}^{1} \overline{4}{}^{\overline{1}} 2{}^{\overline{1}} m{}^{\infty m} 1$    &   {\color{blue} $[E||\overline{4}^{+}_{001}],~[C_{2}||2_{100}],~[C_{2}||m_{\overline{1}10}]$},~$[E||E],~[C_{2}||2_{010}],~[E||2_{001}],~[C_{2}||m_{110}],~[E||\overline{4}^{-}_{001}]$ 
\tabularnewline

34&{\color{red} ${}^{\overline{1}} \overline{4}{}^{\overline{1}} 2{}^{1} m{}^{\infty m} 1$}    &   {\color{blue} $[C_{2}||\overline{4}^{+}_{001}],~[C_{2}||2_{100}],~[E||m_{110}]$},~$[E||E],~[C_{2}||2_{010}],~[E||2_{001}],~[E||m_{\overline{1}10}],~[C_{2}||\overline{4}^{-}_{001}]$ 
\tabularnewline

35& ${}^{\overline{1}} \overline{4}{}^{1} 2{}^{\overline{1}} m{}^{\infty m} 1$    &   {\color{blue} $[C_{2}||\overline{4}^{+}_{001}],~[E||2_{100}],~[C_{2}||m_{110}]$},~$[E||E],~[E||2_{010}],~[E||2_{001}],~[C_{2}||m_{\overline{1}10}],~[C_{2}||\overline{4}^{-}_{001}]$ 
\tabularnewline

36& ${}^{\overline{1}} 4/{}^{\overline{1}} m{}^{\overline{1}} m{}^{1} m{}^{\infty m} 1$    &   {\color{blue} $[C_{2}||4^{+}_{001}],~[C_{2}||\mathcal{P}],~[C_{2}||m_{100}],~[E||m_{\overline{1}10}]$},~$[E||E],~[C_{2}||m_{001}],~[E||2_{100}],~[E||2_{010}],~[E||2_{001}],~$\tabularnewline 
&  &   $[C_{2}||2_{\overline{1}10}],~[E||4_{001}^{-}],~[E||2_{110}],~[C_{2}||m_{010}],~[E||m_{110}],~[E||\overline{4}^+_{001}],~[E||\overline{4}_{001}^{-}]$ 
\tabularnewline
\end{tabular}
\end{ruledtabular}
\label{table2DCSPG2}
\end{table}

\begin{table}[H]
	\renewcommand{\arraystretch}{1.6}
	
	\begin{ruledtabular} %
		\begin{tabular}{lll}
			No. &	Notation & Operators  in $\mathcal G_0$ [$\mathcal G=(\mathcal G_0+[C_2T||E]\mathcal G_0)\ltimes SO(2)$] \tabularnewline
			\hline
			
			56& 			${}^{1} \overline{6}{}^{\overline{1}} m{}^{\overline{1}} 2{}^{\infty m} 1$    &   {\color{blue} $[E||\overline{6}^{+}_{001}],~[C_{2}||m_{100}],~[C_{2}||2_{1\overline{1}0}]$},~$[E||E],~[E||3_{001}^{+}],~[E||3_{001}^{-}],~[C_{2}||2_{120}],~[C_{2}||2_{210}],~[E||m_{001}],~$\tabularnewline 
			& 			 &   $[E||\overline{6}_{001}^{-}],~[C_{2}||m_{110}],~[C_{2}||m_{010}]$ 
			\tabularnewline
			
			57& 		{\color{red}	${}^{\overline{1}} \overline{6}{}^{1} m{}^{\overline{1}} 2{}^{\infty m} 1$ }   &   {\color{blue} $[C_{2}||\overline{6}^{+}_{001}],~[E||m_{100}],~[C_{2}||2_{1\overline{1}0}]$},~$[E||E],~[E||3_{001}^{+}],~[E||3_{001}^{-}],~[C_{2}||2_{120}],~[C_{2}||2_{210}],~[C_{2}||m_{001}],~$\tabularnewline 
			& 			 &   $[C_{2}||\overline{6}_{001}^{-}],~[E||m_{110}],~[E||m_{010}]$ 
			\tabularnewline
			
			58& 			${}^{\overline{1}} \overline{6}{}^{\overline{1}} m{}^{1} 2{}^{\infty m} 1$    &   {\color{blue} $[C_{2}||\overline{6}^{+}_{001}],~[C_{2}||m_{100}],~[E||2_{1\overline{1}0}]$},~$[E||E],~[E||3_{001}^{+}],~[E||3_{001}^{-}],~[E||2_{120}],~[E||2_{210}],~[C_{2}||m_{001}],~$\tabularnewline 
			& 			 &   $[C_{2}||\overline{6}_{001}^{-}],~[C_{2}||m_{110}],~[C_{2}||m_{010}]$ 
			\tabularnewline

59& 			${}^{\overline{1}} 6/{}^{\overline{1}} m{}^{1} m{}^{\overline{1}} m{}^{\infty m} 1$    &   {\color{blue} $[C_{2}||6^{+}_{001}],~[E||\mathcal{P}],~[E||m_{100}],~[C_{2}||m_{1\overline{1}0}]$},~$[E||E],~[E||3_{001}^{+}],~[C_{2}||2_{001}],~[E||3_{001}^{-}],~[C_{2}||6_{001}^{-}],~$\tabularnewline
\end{tabular}
\end{ruledtabular}
\label{table2DCSPG4}
\end{table}

\begin{table}[H]
	\renewcommand{\arraystretch}{1.6}
	\caption{31 collinear spin layer-point group $\mathcal G$: $\mathcal G=(\mathcal G_0+[C_2T||E]\mathcal G_0)\ltimes SO(2)$ for FM.  Only nontrivial operators in $\mathcal G_0$  are listed. The generators of $\mathcal G_0$ are marked in blue.  The subscript $i$ in the second column ($[...]_i$) indicates the $i$-th collinear spin layer-point group obtained from a same 3D collinear spin point group. }
	\begin{ruledtabular} %
		\begin{tabular}{lll}
			No. &	Notation &  Operators  in $\mathcal G_0$ [$\mathcal G=(\mathcal G_0+[C_2T||E]\mathcal G_0)\ltimes SO(2)$]   \tabularnewline
			\hline
1& 		${}^{ \infty m} 1$    &   {\color{blue} $[E||E]$}
\tabularnewline

2& ${}^{1} \overline{1}{}^{ \infty m} 1$    &   {\color{blue} $[E||\mathcal{P}]$},~$[E||E]$
\tabularnewline

3& $[{}^{1} 2{}^{ \infty m} 1]_1$    &   {\color{blue} $[E||2_{001}]$},~$[E||E]$
\tabularnewline

4& $[{}^{1} 2{}^{ \infty m} 1]_2$    &   {\color{blue} $[E||2_{100}]$},~$[E||E]$
\tabularnewline

5& $[{}^{1} m{}^{ \infty m} 1]_1$    &   {\color{blue} $[E||m_{001}]$},~$[E||E]$
\tabularnewline

6& $[{}^{1} m{}^{ \infty m} 1]_2$    &   {\color{blue} $[E||m_{100}]$},~$[E||E]$
\tabularnewline

7& $[{}^{1} 2/{}^{1} m{}^{ \infty m} 1]_1$    &   {\color{blue} $[E||2_{001}],~[E||\mathcal{P}]$},~$[E||E],~[E||m_{001}]$
\tabularnewline
\end{tabular}
	\end{ruledtabular}
	\label{table2DCSPG3}
\end{table}

\begin{table}[H]
	\renewcommand{\arraystretch}{1.6}
	
	\begin{ruledtabular} %
		\begin{tabular}{lll}
			No. &	Notation & Operators  in $\mathcal G_0$ [$\mathcal G=(\mathcal G_0+[C_2T||E]\mathcal G_0)\ltimes SO(2)$] \tabularnewline
			\hline
			
29& ${}^{1} 6{}^{1} m{}^{1} m{}^{ \infty m} 1$    &   {\color{blue} $[E||6^{+}_{001}],~[E||m_{100}],~[E||m_{1\overline{1}0}]$},~$[E||E],~[E||3_{001}^{+}],~[E||2_{001}],~[E||3_{001}^{-}],~[E||6_{001}^{-}],~[E||m_{110}],~$  \tabularnewline
&  &    $[E||m_{010}],~[E||m_{210}],~[E||m_{120}]$
\tabularnewline

30& ${}^{1} \overline{6}{}^{1} m{}^{1} 2{}^{ \infty m} 1$    &   {\color{blue} $[E||\overline{6}^{+}_{001}],~[E||m_{100}],~[E||2_{1\overline{1}0}]$},~$[E||E],~[E||3_{001}^{+}],~[E||2_{120}],~[E||3_{001}^{-}],~[E||\overline{6}_{001}^{-}],~[E||2_{210}],~$  \tabularnewline
&  &    $[E||m_{001}],~[E||m_{110}],~[E||m_{010}]$
\tabularnewline

31& ${}^{1} 6/{}^{1} m{}^{1} m{}^{1} m{}^{ \infty m} 1$    &   {\color{blue} $[E||6^{+}_{001}],~[E||\mathcal{P}],~[E||m_{100}],~[E||m_{1\overline{1}0}]$},~$[E||E],~[E||3_{001}^{+}],~[E||2_{001}],~[E||3_{001}^{-}],~[E||6_{001}^{-}],~$  \tabularnewline
&  &    $[E||2_{120}],~[E||2_{110}],~[E||2_{210}],~[E||2_{010}],~[E||2_{1\overline{1}0}],~[E||m_{001}],~[E||\overline{6}_{001}^+],~[E||\overline{3}_{001}^{+}],~[E||\overline{3}_{001}^{-}],~$  \tabularnewline
&  &    $[E||\overline{6}_{001}^{-}],~[E||m_{110}],~[E||m_{120}],~[E||m_{010}],~[E||m_{210}],~[E||2_{100}]$
\tabularnewline

		\end{tabular}
	\end{ruledtabular}
\end{table}

\subsection{Operators of 90 3D collinear spin point groups} \label{3DSPG}

\begin{table}[H]
   \renewcommand{\arraystretch}{1.6}
\caption{58 collinear spin point group $\mathcal G$: $\mathcal G=(\mathcal G_0+[C_2T||E]\mathcal G_0)\ltimes SO(2)$ for AFM.  Only nontrivial operators in $\mathcal G_0$  are listed. The generators of $\mathcal G_0$ are marked in blue. The groups allowing for 3D spin-AFE are highlighted by red.}
\begin{ruledtabular} %
	\begin{tabular}{lll}
		No. &	Notation & Operators  in $\mathcal G_0$ [$\mathcal G=(\mathcal G_0+[C_2T||E]\mathcal G_0)\ltimes SO(2)$]  \tabularnewline
		\hline
1&  {\color{red}${}^{\overline{1}} \overline{1}{}^{\infty m} 1$}    &   {\color{blue} $[C_{2}||\mathcal{P}]$},~$[E||E]$ 
\tabularnewline

2& ${}^{\overline{1}} 2{}^{\infty m} 1$    &   {\color{blue} $[C_{2}||2_{010}]$},~$[E||E]$ 
\tabularnewline

3& ${}^{\overline{1}} m{}^{\infty m} 1$    &   {\color{blue} $[C_{2}||m_{010}]$},~$[E||E]$ 
\tabularnewline

4& {\color{red} ${}^{1} 2/{}^{\overline{1}} m{}^{\infty m} 1$ }   &   {\color{blue} $[E||2_{010}],~[C_{2}||\mathcal{P}]$},~$[E||E],~[C_{2}||m_{010}]$ 
\tabularnewline

5& {\color{red}${}^{\overline{1}} 2/{}^{1} m{}^{\infty m} 1$ }   &   {\color{blue} $[C_{2}||\mathcal{P}],~[E||m_{010}]$},~$[E||E],~[C_{2}||2_{010}]$ 
\tabularnewline

6& ${}^{\overline{1}} 2/{}^{\overline{1}} m{}^{\infty m} 1$    &   {\color{blue} $[C_{2}||2_{010}],~[E||\mathcal{P}]$},~$[E||E],~[C_{2}||m_{010}]$ 
\tabularnewline

7& ${}^{\overline{1}} m{}^{\overline{1}} m{}^{1} 2{}^{\infty m} 1$    &   {\color{blue} $[C_{2}||m_{100}],~[E||2_{001}]$},~$[E||E],~[C_{2}||m_{010}]$ 
\tabularnewline
\end{tabular}
	\end{ruledtabular}
	\label{table3DCSPG1}
\end{table}

\begin{table}[H]
   \renewcommand{\arraystretch}{1.6}

	\begin{ruledtabular} %
		\begin{tabular}{lll}
				No. &	Notation & Operators  in $\mathcal G_0$ [$\mathcal G=(\mathcal G_0+[C_2T||E]\mathcal G_0)\ltimes SO(2)$]   \tabularnewline
\hline

30& {\color{red}${}^{\overline{1}} \overline{3}{}^{\infty m} 1$  }  &   {\color{blue} $[C_{2}||\overline{3}^{+}_{001}]$},~$[E||E],~[E||3_{001}^{+}],~[E||3_{001}^{-}],~[C_{2}||\overline{3}_{001}^{-}],~[C_{2}||\mathcal{P}]$ 
\tabularnewline

31& {\color{red}${}^{1} 3{}^{\overline{1}} 2{}^{\infty m} 1$  }  &   {\color{blue} $[E||3^{+}_{001}],~[C_{2}||2_{100}]$},~$[E||E],~[E||3_{001}^{-}],~[C_{2}||2_{110}],~[C_{2}||2_{010}]$ 
\tabularnewline

32& ${}^{1} 3{}^{\overline{1}} m{}^{\infty m} 1$    &   {\color{blue} $[E||3^{+}_{001}],~[C_{2}||m_{100}]$},~$[E||E],~[E||3_{001}^{-}],~[C_{2}||m_{110}],~[C_{2}||m_{010}]$ 
\tabularnewline

33& ${}^{1} \overline{3}{}^{\overline{1}} m{}^{\infty m} 1$    &   {\color{blue} $[E||\overline{3}^{+}_{001}],~[C_{2}||m_{100}]$},~$[E||E],~[E||3^{+}_{001}],~[E||3^{-}_{001}],~[C_{2}||2_{100}],~[C_{2}||2_{110}],~[C_{2}||2_{010}],~[E||\mathcal{P}],~$\tabularnewline 
&  &   $[E||\overline{3}_{001}^{-}],~[C_{2}||m_{110}],~[C_{2}||m_{010}]$ 
\tabularnewline

34& {\color{red}${}^{\overline{1}} \overline{3}{}^{1} m{}^{\infty m} 1$ }   &   {\color{blue} $[C_{2}||\overline{3}^{+}_{001}],~[E||m_{100}]$},~$[E||E],~[E||3^{+}_{001}],~[E||3^{-}_{001}],~[C_{2}||2_{100}],~[C_{2}||2_{110}],~[C_{2}||2_{010}],~[C_{2}||\mathcal{P}],~$\tabularnewline 
&  &   $[C_{2}||\overline{3}_{001}^{-}],~[E||m_{110}],~[E||m_{010}]$ 
\tabularnewline
\end{tabular}
	\end{ruledtabular}
	\label{table3DCSPG2}
\end{table}

\begin{table}[H]
	\renewcommand{\arraystretch}{1.6}

	\begin{ruledtabular} %
		\begin{tabular}{lll}
				No. &	Notation & Operators  in $\mathcal G_0$ [$\mathcal G=(\mathcal G_0+[C_2T||E]\mathcal G_0)\ltimes SO(2)$]  \tabularnewline
			\hline

50& 			${}^{1} 6/{}^{1} m{}^{\overline{1}} m{}^{\overline{1}} m{}^{\infty m} 1$    &   {\color{blue} $[E||6^{+}_{001}],~[E||\mathcal{P}],~[C_{2}||m_{100}],~[C_{2}||m_{1\overline{1}0}]$},~$[E||E],~[E||3_{001}^{+}],~[E||2_{001}],~[E||3_{001}^{-}],~[E||6_{001}^{-}],~$\tabularnewline 
& 			 &   $[C_{2}||2_{1\overline{1}0}],~[C_{2}||2_{110}],~[C_{2}||2_{120}],~[C_{2}||2_{010}],~[C_{2}||2_{210}],~[E||m_{001}],~[E||\overline{6}^+_{001}],~[E||\overline{3}^+_{001}],~$\tabularnewline 
& 			 &   $[E||\overline{3}_{001}^{-}],~[E||\overline{6}_{001}^{-}],~[C_{2}||m_{110}],~[C_{2}||m_{120}],~[C_{2}||m_{010}],~[C_{2}||m_{210}],~[C_{2}||2_{100}]$\tabularnewline 

51& 		{\color{red}	${}^{1} 6/{}^{\overline{1}} m{}^{1} m{}^{1} m{}^{\infty m} 1$  }  &   {\color{blue} $[E||6^{+}_{001}],~[C_{2}||\mathcal{P}],~[E||m_{100}],~[E||m_{1\overline{1}0}]$},~$[E||E],~[E||3_{001}^{+}],~[E||2_{001}],~[E||3_{001}^{-}],~[E||6_{001}^{-}],~$\tabularnewline 
& 			 &   $[C_{2}||2_{1\overline{1}0}],~[C_{2}||2_{110}],~[C_{2}||2_{120}],~[C_{2}||2_{010}],~[C_{2}||2_{210}],~[C_{2}||m_{001}],~[C_{2}||\overline{6}^+_{001}],~[C_{2}||\overline{3}_{001}^{+}],~$\tabularnewline 
& 			 &   $[C_{2}||\overline{3}_{001}^{-}],~[C_{2}||\overline{6}_{001}^{-}],~[E||m_{110}],~[E||m_{120}],~[E||m_{010}],~[E||m_{210}],~[C_{2}||2_{100}]$ 
\tabularnewline

52& 			${}^{1} 6/{}^{\overline{1}} m{}^{\overline{1}} m{}^{\overline{1}} m{}^{\infty m} 1$    &   {\color{blue} $[E||6^{+}_{001}],~[C_{2}||\mathcal{P}],~[C_{2}||m_{100}],~[C_{2}||m_{1\overline{1}0}]$},~$[E||E],~[E||3_{001}^{+}],~[E||2_{001}],~[E||3_{001}^{-}],~[E||6_{001}^{-}],~$\tabularnewline
\end{tabular}
	\end{ruledtabular}
	\label{table3DCSPG4}
\end{table}

\begin{table}[H]
   \renewcommand{\arraystretch}{1.6}
\caption{32 collinear spin point group $\mathcal G$: $\mathcal G=(\mathcal G_0+[C_2T||E]\mathcal G_0)\ltimes SO(2)$ for FM.  Only nontrivial operators in $\mathcal G_0$  are listed. The generators of $\mathcal G_0$ are marked in blue. }
\begin{ruledtabular} %
	\begin{tabular}{lll}
			No. &	Notation & Operators  in $\mathcal G_0$ [$\mathcal G=(\mathcal G_0+[C_2T||E]\mathcal G_0)\ltimes SO(2)$] \tabularnewline
		\hline
1& 		${}^{ \infty m} 1$    &   {\color{blue} $[E||E]$}
\tabularnewline

2& ${}^{1} \overline{1}{}^{ \infty m} 1$    &   {\color{blue} $[E||\mathcal{P}]$},~$[E||E]$
\tabularnewline

3& ${}^{1} 2{}^{ \infty m} 1$    &   {\color{blue} $[E||2_{010}]$},~$[E||E]$
\tabularnewline

4& ${}^{1} m{}^{ \infty m} 1$    &   {\color{blue} $[E||m_{010}]$},~$[E||E]$
\tabularnewline

5& ${}^{1} 2/{}^{1} m{}^{ \infty m} 1$    &   {\color{blue} $[E||2_{010}],~[E||\mathcal{P}]$},~$[E||E],~[E||m_{010}]$
\tabularnewline

6& ${}^{1} m{}^{1} m{}^{1} 2{}^{ \infty m} 1$    &   {\color{blue} $[E||m_{100}],~[E||2_{001}]$},~$[E||E],~[E||m_{010}]$
\tabularnewline

7& ${}^{1} 2{}^{1} 2{}^{1} 2{}^{ \infty m} 1$    &   {\color{blue} $[E||2_{010}],~[E||2_{100}]$},~$[E||E],~[E||2_{001}]$
\tabularnewline
\end{tabular}
	\end{ruledtabular}
	\label{table3DCSPG3}
\end{table}

\begin{table}[H]
	\renewcommand{\arraystretch}{1.6}
	\begin{ruledtabular} %
		\begin{tabular}{lll}
			No. &	Notation & Operators  in $\mathcal G_0$ [$\mathcal G=(\mathcal G_0+[C_2T||E]\mathcal G_0)\ltimes SO(2)$]    \tabularnewline
			\hline
			
			28& ${}^{1} 2{}^{1} 3{}^{m} 1$    &  {\color{blue}  $[E||2_{001}],~[E||3^{+}_{111}]$},~$[E||E],~[E||2_{010}],~[E||2_{100}],~[E||3_{\overline{1}1\overline{1}}^{+}],~[E||3_{1\overline{1}\overline{1}}^{+}],~[E||3_{\overline{1}\overline{1}1}^{+}],~[E||3_{111}^{-}],~$  \tabularnewline
			&  &    $[E||3_{1\overline{1}\overline{1}}^{-}],~[E||3_{\overline{1}\overline{1}1}^{-}],~[E||3_{\overline{1}1\overline{1}}^{-}]$
			\tabularnewline
			
			29& ${}^{1} 2/{}^{1} m{}^{1} \overline{3}{}^{ m} 1$    &  
			{\color{blue}  $[E||2_{001}],~[E||m_{001}],~[E||\overline{3}^{+}_{111}]$},~$[E||E],~[E||2_{100}],~[E||2_{010}],~[E||3_{111}^{-}],~[E||3_{11\overline{1}}^{+}],~[E||3_{\overline{1}11}^{+}],~$  \tabularnewline
			&  &    $[E||3_{1\overline{1}1}^{+}],~[E||3_{111}^{+}],~[E||3_{1\overline{1}1}^{-}],~[E||3_{11\overline{1}}^{-}],~[E||3_{\overline{1}11}^{-}],~[E||\mathcal{P}],~[E||m_{100}],~[E||m_{010}],~[E||\overline{3}_{111}^{-}],~$  \tabularnewline
			&  &    $[E||\overline{3}_{11\overline{1}}^{+}],~[E||\overline{3}_{\overline{1}11}^{+}],~[E||\overline{3}_{1\overline{1}1}^{+}],~[E||\overline{3}_{1\overline{1}1}^{-}],~[E||\overline{3}_{11\overline{1}}^{-}],~[E||\overline{3}_{\overline{1}11}^{-}]$
			\tabularnewline
			
			30& ${}^{1} \overline{4}{}^{1} 3{}^{1} m{}^{ m} 1$    &  
			{\color{blue}  $[E||\overline{4}^{+}_{001}],~[E||3^{+}_{111}],~[E||m_{110}]$},~$[E||E],~[E||2_{100}],~[E||2_{010}],~[E||2_{001}],~[E||3_{111}^{-}],~[E||3_{11\overline{1}}^{+}],~[E||3_{\overline{1}11}^{+}],~$  \tabularnewline
			&  &    $[E||3_{1\overline{1}1}^{+}],~[E||\overline{4}_{010}],~[E||3_{1\overline{1}1}^{-}],~[E||3_{11\overline{1}}^{-}],~[E||3_{\overline{1}11}^{-}],~[E||m_{\overline{1}10}],~[E||m_{101}],~[E||\overline{4}_{001}^{-}],~$  \tabularnewline
\end{tabular}
	\end{ruledtabular}
	\label{table3DCSPG5}
\end{table}

\newpage

\section{Operators of 125 2D magnetic layer-point groups }\label{MPG}

\begin{table}[H]
	\renewcommand{\arraystretch}{1.5}
	\caption{125 magnetic layer-point groups.  Shubnikov (Litvin) notation is presented in  second column. The generators for each group are  marked in blue.
 The groups allowing for mirror-AFE are highlighted by red. The superscript ($'$) in  third column indicates the presence of $T$ symmetry. For example, $E'$ denotes the operator of $ET$.}
	\begin{ruledtabular} %
		\begin{tabular}{@{}l@{\hspace{0mm}}l@{\hspace{0mm}}l}
			\noindent
		No. &	Notation & Operators    \tabularnewline
			\hline
1 & 			$1.1\,(1.1.1)$    &    ${\color{blue}  E}$   \tabularnewline 

2& 			$1.1^{\prime}\,(1.2.2)$    &    ${\color{blue}  E^{\prime}},~E$   \tabularnewline 

3& 			$\overline{1}.1\,(2.1.3)$    &    ${\color{blue}  \mathcal{P}},~E$   \tabularnewline	 

4& 			$\overline{1}.1^{\prime}\,(2.2.4)$    &    ${\color{blue}  \mathcal{P},~E^{\prime}},~E,~\mathcal{P}^{\prime}$   \tabularnewline	 

5& 			$\overline{1}^{\prime}\,(2.3.5)$    &    ${\color{blue}  \mathcal{P}^{\prime}},~E$   \tabularnewline 

6& 			$2.1\,(3.1.6)$    &    ${\color{blue}  2_{001}},~E$   \tabularnewline 

7& 		$2.1\,(3.1.6)$	    &    ${\color{blue}  2_{100}},~E$   \tabularnewline
\end{tabular}
	\end{ruledtabular}
	\label{table2DMPG1}
\end{table}

\begin{table}[H]
	\renewcommand{\arraystretch}{1.5}

	\begin{ruledtabular} %
		\begin{tabular}{lll}
			No. &	Notation & Operators   \tabularnewline
			\hline

			36& 			$m^{\prime}m2^{\prime}\,(7.3.22)$    &    ${\color{blue}  m_{010},~2_{001}^{\prime}},~E,~m_{100}^{\prime}$   \tabularnewline 
			
			37& 			$m^{\prime}m2^{\prime}\,(7.3.22)$    &    ${\color{blue}  m_{001},~2_{010}^{\prime}},~E,~m_{100}^{\prime}$   \tabularnewline
			
						38& 			$m^{\prime}m2^{\prime}\,(7.3.22)$    &    ${\color{blue}  m_{100},~2_{010}^{\prime}},~E,~m_{001}^{\prime}$   \tabularnewline
			
			39& 			$m^{\prime}m^{\prime}2\,(7.4.23)$    &    ${\color{blue}  2_{001},~m_{100}^{\prime}},~E,~m_{010}^{\prime}$   \tabularnewline 
			
			40& 			$m^{\prime}m^{\prime}2\,(7.4.23)$    &    ${\color{blue}  2_{010},~m_{100}^{\prime}},~E,~m_{001}^{\prime}$   \tabularnewline 
			
			41& 			$mmm.1\,(8.1.24)$    &    ${\color{blue}  2_{100},~2_{010},~\mathcal{P}},~E,~2_{001},~m_{100},~m_{010},~m_{001}$   \tabularnewline 
			
			42& 			$mmm.1^{\prime}\,(8.2.25)$    &    ${\color{blue}  2_{100},~2_{010},~\mathcal{P},~E^{\prime}},~E,~2_{001},~m_{100},~m_{010},~m_{001},~2_{100}^{\prime},~2_{010}^{\prime},~2_{001}^{\prime},~\mathcal{P}^{\prime},~m_{100}^{\prime},~m_{010}^{\prime},~m_{001}^{\prime}$   \tabularnewline
\end{tabular}
	\end{ruledtabular}
	\label{table2DMPG2}
\end{table}

\begin{table}[H]
	\renewcommand{\arraystretch}{1.6}

	\begin{ruledtabular} %
		\begin{tabular}{lll}
			No. &	Notation & Operators    \tabularnewline
			\hline

			74& 			$4/m^{\prime}mm\,(15.3.55)$    &    ${\color{blue}  4_{001}^{+},~m_{001}^{\prime},~m_{100},~m_{110}},~E,~2_{100}^{\prime},~2_{010}^{\prime},~2_{001},~2_{\overline{1}10}^{\prime},~4_{001}^{- },~2_{110}^{\prime},~\mathcal{P}^{\prime},~m_{010},~m_{\overline{1}10},~\overline{4}_{001}^{+ \prime},~\overline{4}_{001}^{- \prime}$   \tabularnewline

			75& 			$4^{\prime}/mm^{\prime}m\,(15.4.56)$    &    ${\color{blue}  4_{001}^{+ \prime},~m_{001},~m_{100}^{\prime},~m_{110}},~E,~2_{100}^{\prime},~2_{010}^{\prime},~2_{001},~2_{\overline{1}10},~4_{001}^{- \prime},~2_{110},~\mathcal{P},~m_{010}^{\prime},~m_{\overline{1}10},~\overline{4}_{001}^{+ \prime},~\overline{4}_{001}^{- \prime}$   \tabularnewline 
			
			76& 			$4^{\prime}/m^{\prime}m^{\prime}m\,(15.5.57)$    &    ${\color{blue}  4_{001}^{+ \prime},~m_{001}^{\prime},~m_{100}^{\prime},~m_{110}},~E,~2_{100},~2_{010},~2_{001},~2_{\overline{1}10}^{\prime},~4_{001}^{- },~2_{110}^{\prime},~\mathcal{P}^{\prime},~m_{010}^{\prime},~m_{\overline{1}10},~\overline{4}_{001}^+,~\overline{4}_{001}^{-}$   \tabularnewline 
			
			77& 			$4/mm^{\prime}m^{\prime}\,(15.6.58)$    &    ${\color{blue}  4_{001}^{+},~m_{001},~m_{100}^{\prime},~m_{110}^{\prime}},~E,~2_{100}^{\prime},~2_{010}^{\prime},~2_{001},~2_{\overline{1}10}^{\prime},~4_{001}^{- },~2_{110}^{\prime},~\mathcal{P},~m_{010}^{\prime},~m_{\overline{1}10}^{\prime},~\overline{4}_{001}^+,~\overline{4}_{001}^{-}$   \tabularnewline 
			
			78& 			$4/m^{\prime}m^{\prime}m^{\prime}\,(15.7.59)$    &    ${\color{blue}  4_{001}^{+},~m_{001}^{\prime},~m_{100}^{\prime},~m_{110}^{\prime}},~E,~2_{100},~2_{010},~2_{001},~2_{\overline{1}10},~4_{001}^{- },~2_{110},~\mathcal{P}^{\prime},~m_{010}^{\prime},~m_{\overline{1}10}^{\prime},~\overline{4}_{001}^{+ \prime},~\overline{4}_{001}^{- \prime}$   \tabularnewline 
			
			79& 			$3.1\,(16.1.60)$    &    ${\color{blue}  3_{001}^+},~E,~3_{001}^{-}$   \tabularnewline
\end{tabular}
	\end{ruledtabular}
	\label{table2DMPG3}
\end{table}

\begin{table}[H]
	\renewcommand{\arraystretch}{1.6}
	
	\begin{ruledtabular} %
		\begin{tabular}{lll}
			No. &	Notation & Operators    \tabularnewline
			\hline
			
			106& 			$622.1\,(24.1.87)$    &    ${\color{blue}  6_{001}^+,~2_{100},~2_{210}},~E,~3_{001}^{+},~2_{001},~3_{001}^{-},~6_{001}^{-},~2_{110},~2_{120},~2_{010},~2_{1\overline{1}0}$   \tabularnewline 
			
			107& 			$622.1^{\prime}\,(24.2.88)$    &    ${\color{blue}  6_{001}^+,~2_{100},~2_{210},~E^{\prime}},~E,~3_{001}^{+},~2_{001},~3_{001}^{-},~6_{001}^{-},~2_{110},~2_{120},~2_{010},~2_{1\overline{1}0},~ 
			6_{001}^{+ \prime},~2_{100}^{\prime},~2_{210}^{\prime},~3_{001}^{+ \prime},~2_{001}^{\prime},~$   \tabularnewline
			& & $3_{001}^{- \prime},~6_{001}^{- \prime},~2_{110}^{\prime},~2_{120}^{\prime},~2_{010}^{\prime},~2_{1\overline{1}0}^{\prime}$  \tabularnewline
			
			108& 			$6^{\prime}22^{\prime}\,(24.3.89)$    &    ${\color{blue}  6_{001}^{+ \prime},~2_{100},~2_{210}^{\prime}},~E,~3_{001}^{+},~2_{001}^{\prime},~3_{001}^{-},~6_{001}^{- \prime},~2_{110},~2_{120}^{\prime},~2_{010},~2_{1\overline{1}0}^{\prime}$   \tabularnewline 
			
			109& 			$62^{\prime}2^{\prime}\,(24.4.90)$    &    ${\color{blue}  6_{001}^+,~2_{100}^{\prime},~2_{210}^{\prime}},~E,~3_{001}^{+},~2_{001},~3_{001}^{-},~6_{001}^{-},~2_{110}^{\prime},~2_{120}^{\prime},~2_{010}^{\prime},~2_{1\overline{1}0}^{\prime}$   \tabularnewline 
			
			110& 			$6mm.1\,(25.1.91)$    &    ${\color{blue}  6_{001}^+,~m_{100},~m_{210}},~E,~3_{001}^{+},~2_{001},~3_{001}^{-},~6_{001}^{-},~m_{110},~m_{120},~m_{010},~m_{1\overline{1}0}$   \tabularnewline 
			
			111& 			$6mm.1^{\prime}\,(25.2.92)$    &    ${\color{blue}  6_{001}^+,~m_{100},~m_{210},~E^{\prime}},~E,~3_{001}^{+},~2_{001},~3_{001}^{-},~6_{001}^{-},~m_{110},~m_{120},~m_{010},~m_{1\overline{1}0},~ 
			6_{001}^{+ \prime},~m_{100}^{\prime},~m_{210}^{\prime},~3_{001}^{+ \prime},~$   \tabularnewline
\end{tabular}
	\end{ruledtabular}
	\label{table2DMPG4}
\end{table}

\bibliography{supref}